%% file: main.tex
\documentclass[11pt,letterpaper]{article}
\usepackage[margin=1in]{geometry}

\usepackage{amsmath}
\usepackage{amssymb}
\usepackage{amsthm}
\usepackage{mathtools}
\usepackage[numbers,sort&compress]{natbib}
\usepackage{hyperref}
\usepackage{cleveref}
\usepackage{comment}
\usepackage{array}
\usepackage{thmtools}
\usepackage{thm-restate}

\usepackage{algorithm}
\usepackage{algpseudocode}
\usepackage{graphicx}
\usepackage{threeparttable}

\crefname{claim}{claim}{claims}
\Crefname{claim}{Claim}{Claims}
\crefname{observation}{observation}{observations}
\Crefname{observation}{Observation}{Observations}
\crefname{problem}{problem}{problem}
\Crefname{problem}{Problem}{Problem}
\crefname{algstep}{step}{steps}
\Crefname{algstep}{Step}{Steps}
\newcommand{\steplabel}[1]{\label[algstep]{#1}}

\newtheorem{problem}{Problem}
\newtheorem{theorem}{Theorem}
\newtheorem{lemma}{Lemma}

\newtheorem{observation}{Observation}
\newtheorem{hypothesis}{Hypothesis}
\crefname{hypothesis}{hypothesis}{hypotheses}
\Crefname{hypothesis}{Hypothesis}{Hypotheses}
\usepackage{xcolor}
\newcommand{\EMPH}[1]{\textcolor{red}{#1}}

\newcommand{\hector}[1]{{\color{blue}[Hector: #1]}}

\title{Reachability Diameter: Beyond the Square-Root Barrier, with Polynomial Hardness}

\author{
    Hector Tierno\thanks{Both authors are supported by NSF grant
    CCF-2517033 and NSF CAREER Award CCF-2237288.}\\
    University of Massachusetts\\
    \texttt{htierno@umass.edu}
    \and
    Shuang Yang\\
    University of Massachusetts\\
    \texttt{shuangyang@umass.edu}
}

\date{}

\begin{document}
\maketitle

\begin{abstract}
    The reachability diameter of a directed graph is the maximum finite shortest-path distance across all pairs of vertices.
    It has received recent attention under this name \cite{haeupler2026reducing, bals2026diameter},
    although the underlying quantity has long played a central role in shortcut set and parallel reachability. 
    Notably Bals et al. \cite{bals2026diameter} establish hardness for every polynomial-factor approximation in nearly-linear time for weighted digraphs.
    Yet they give $n^{1/2+o(1)}$-approximation in the unweighted regime with nearly-linear running time,
    and their conditional lower bounds rule out certain constants only,
    leaving open whether polynomial loss is necessary.

    We give a randomized algorithm that estimates the reachability diameter of an unweighted digraph within a factor $n^{1/3+o(1)}$, 
    with high probability, 
    in $\widetilde{O}(m+n)$ time.
    We complement our algorithm with a lower bound conditional on the 3-uniform Hyperclique hypothesis,
    ruling out an $n^{2/15-\varepsilon}$-approximation in
    $\widetilde{O}(m+n)$ time for every fixed
    $\varepsilon\in(0,2/15)$,
    establishing the necessity of polynomial loss for nearly-linear time algorithms.
\end{abstract}

\input{Introduction.tex}
\input{Overview}
\input{LowerBound.tex}
\input{UpperBound.tex}

\bibliographystyle{ACM-Reference-Format}
\bibliography{references}

\end{document}

%% file: Introduction.tex
\section{Introduction}

The reachability diameter of a graph is the maximum $u$-to-$v$ shortest-path distance across all pairs $u, v$ such that $v$ is reachable from $u$.
It coincides with the classic notion of diameter whenever every vertex can reach every other vertex.
Otherwise, the classical diameter
is infinite, potentially obscuring the distances between pairs
that remain reachable.
For example, adding a single isolated
vertex to a connected undirected graph makes its classical
diameter infinite without changing any distances between the
original vertices.
Similarly, deleting one edge of a directed
cycle produces a directed path: its classical diameter becomes
infinite, but its reachability diameter remains unchanged.
Thus, 
reachability diameter generalizes classical diameter in applications where finite distances are more informative than infinite ones.
The reachability diameter is also the relevant quantity in the shortcut set problem, 
where the goal is to efficiently construct a sparse set of reachability-preserving edges, 
which are added to the input digraph to reduce its reachability diameter to a prescribed \emph{hopbound} (each edge a \emph{hop}).
Despite this, it has received much less attention as a computational problem in its own right. 

Bals, Blikstad, Dadush, Nazari, and Schmidt \cite{bals2026diameter} pioneered the study of efficiently estimating this quantity in unweighted digraphs.
Throughout, $n$ and $m$ denote respectively the numbers of vertices and edges, and a $\rho$-approximation to the reachability diameter $D$ is an estimate $\widehat{D}$ satisfying $D/\rho \le \widehat{D} \le D$.
They give an $n^{1/2 + o(1)}$-approximation algorithm running in nearly-linear $\widetilde{O}(m+n)$ time, yet surprisingly their conditional hardness results only rule out approximation factors below certain constants.
Specifically, the Boolean matrix multiplication (BMM) hypothesis rules out combinatorial better-than-$2$ approximations in truly subcubic time, 
while the orthogonal vectors (OV) hypothesis rules out better-than-$3/2$ approximations in truly subquadratic time, even on sparse graphs.
In matrix-multiplication time they show that exact computation is already possible. 
By contrast, 
assuming $\omega > 2$,
the simplicial vertex hypothesis (SV) rules out any polynomial-factor approximation in near-linear time in the weighted regime (\Cref{tab:diameter-landscape}).
We focus on the near-linear time regime, improving both the approximation guarantee and the conditional hardness bound.

\paragraph{The $n^{1/2}$ barrier.} 
The $n^{1/2+o(1)}$-approximation algorithm of Bals et al.~\cite{bals2026diameter}
inherits its approximation factor from the hopbound of the underlying
shortcut algorithm,
which seeks to efficiently match a well known folklore construction achieving a $\widetilde{O}(n^{1/2})$ hopbound.
Improving the approximation factor requires going beyond what this
hopbound directly implies. We recall the construction:
given an input unweighted digraph, sample $\lceil n^{1/2} \rceil$ vertices called \emph{pivots} uniformly at random without replacement.
Run BFS from each pivot $s$ and include the edge $(s, s^\prime)$ for each other pivot $s^\prime$ reachable from $s$.
Fix one shortest path for each ordered pair of vertices with finite distance.
Then every sufficiently long such path contains a pivot near each endpoint with high probability, 
allowing the segment between them to be traversed in a single hop via the associated shortcut edge.
With $\lceil n^{1/2} \rceil$ pivot samples the construction adds only $O(n)$ edges, 
and the reachability diameter of the augmented graph is bounded by $\widetilde{O}(n^{1/2})$.

To see how estimating reachability diameter inherits the folklore $n^{1/2}$ barrier, consider a simple modification of the folklore algorithm: whenever a BFS is initiated at a pivot, record the largest depth seen during the search.
In the trivial case, output 0, otherwise report the maximum of 1 and the recorded depths, and retain the shortcut set constructed along the way for the analysis.
This already yields a $\widetilde{O}(n^{1/2})$-approximation to the reachability diameter.
To see why, condition on the high-probability event described above, 
and let $P$ be a fixed diametral path.
The construction partitions $P$ into at most $\widetilde O(n^{1/2})$ segments, 
each consisting of either a single original edge or a subpath whose endpoints are joined by a shortcut.
By an application of the pigeon-hole principle, 
some segment has length $\widetilde{\Omega}(D/n^{1/2})$.
If this segment corresponds to a shortcut edge $(s, s^\prime)$, 
BFS initiated at $s$ reports the distance exactly.
The limitation of this approach is the cost of search, yielding an $O((m+n)\sqrt n)$ running time, which exceeds our near-linear time target.
Efficient shortcut constructions would offer a route to a comparable approximation
guarantee for estimating reachability diameter in the desired near-linear time.

\paragraph{History of efficient shortcut construction.}
The challenge of constructing sparse shortcut sets efficiently has long been central to parallel reachability, 
the problem of finding all vertices reachable from a given source.
Parallel BFS explores each layer concurrently, which can require as many sequential rounds as its reachability diameter.
Motivated by parallel reachability, 
Thorup \cite{thorup1993shortcutting} formalized the problem of reducing a digraph's reachability diameter by adding a set of reachability-preserving edges.
The construction must be efficient, and the shortcut set sparse, to retain the near-linear total computational cost of BFS.
The first shortcut-set algorithm achieving, with high probability, 
polynomially sublinear $\widetilde{O}(n^{2/3})$ reachability diameter with $\widetilde{O}(m)$ running time and $\widetilde{O}(n)$ size was discovered by Fineman \cite{fineman2018reachability},
refining the sample and search approach of Ullman and Yannakakis \cite{ullman1991transitive}.

Soon after, Jambulapati, Liu, and Sidford \cite{liu2019reachability} obtained the first efficient and sparse construction (JLS algorithm) whose bound $O(n^{1/2+o(1)})$ nearly matches the folklore algorithm's.
For a fixed path $P$, 
instead of a pivot near each endpoint being used to shortcut the middle segment, 
the JLS algorithm partitions the graph, 
which fragments $P$,
and constructs a shortcut for each fragment.
At every recursive level its fragments are either resolved by being shortcut through a sampled pivot, 
or become recursive subproblems, further subdivided and eventually shortcut with later samples.
The expected number of fragments of $P$ over the course of the recursion is at most $n^{1/2+o(1)}$.
It is at this point that shortcut construction meets diameter estimation.
Bals et al. \cite{bals2026diameter} show that the JLS algorithm's partitioning preserves original graph distances in each child,
and by an application of the triangle inequality, 
search from pivots reveals a distance at least half the length of fragments of shortest paths which they resolve. 
Combining these ideas and the prior pigeon-hole argument, they achieve the first $n^{1/2+o(1)}$-approximation reachability diameter algorithm in unweighted digraphs.
By tying the approximation factor to the shortcut hopbound, the approach
of Bals et al. carries the folklore $n^{1/2}$ benchmark into reachability
diameter estimation, making it an apparent barrier for this problem as well.

Further developments in shortcut sets emphasize $n^{1/2}$ as a significant benchmark. 
Kogan and Parter achieve a substantially smaller hopbound of $\widetilde{O}(n^{1/3})$ \cite{koganparter2022shortcuts},
but their construction takes superlinear $\widetilde{O}(mn^{1/3})$ time \cite{kogan2023faster}.
Vikrant Ashvinkumar, Aaron Bernstein, Maximilian Probst Gutenberg, and Thatchaphol Saranurak \cite{ashvinkumar2026parallel} retain near-linear work and improve the reachability diameter to $\left(\frac{n^{\omega+o(1)}}{m}\right)^{1/(2\omega-2)}$ on dense graphs allowing $\widetilde{O}(m)$ shortcut edges,
where $\omega$ is the matrix multiplication exponent. 
However, their guarantee remains $n^{1/2+o(1)}$ on sparse graphs.
This motivates the following refinement of the question of~\cite{bals2026diameter}:
\begin{problem}\label{pbl:BalsOpenPbl}
    Can reachability diameter estimation overcome $n^{1/2}$ in better-than-matrix-multipli\-cation time, without a corresponding improvement in shortcut construction?
\end{problem}
We answer this question in the affirmative, giving a randomized
$\widetilde{O}(m+n)$-time algorithm that achieves an
$n^{1/3+o(1)}$-approximation with high probability.
\begin{theorem}\label{thm:main-theorem}
    There is a randomized algorithm computing a $n^{1/3+o(1)}$-approximation to the reachability diameter of an unweighted digraph with high probability, running in $\widetilde{O}(m+n)$ time.
\end{theorem}
Our algorithm builds on the same recursive partitioning framework of Jambulapati et al. \cite{liu2019reachability} also used by Bals et al. 
Following a fixed diametral path,
at each level of the recursion random samples either further subdivide an existing fragment, 
or resolve it entirely.
Bals et al. show that resolved fragments reveal distances in the base input graph,  
and thus the final count of resolved fragments controls the approximation ratio of their algorithm.
In contrast, 
with careful analysis we show that our approach obtains a measurement earlier in the recursion,
before repeated decompositions have cut the length of the diametral path into too many pieces.
The difficulty lies in detecting a sufficiently long piece when it is unresolved.
To address this, we adapt a randomized procedure of Cohen \cite{cohen1997size} for estimating the size of a neighborhood to finding length-witnesses of unresolved pieces.
Our analysis balances bounds of Ashvinkumar et al. \cite{ashvinkumar2026parallel} on the number of large recursive subproblems,
with the success rate of the adapted procedure,
breaking the $n^{1/2}$ approximation barrier for reachability diameter.
The algorithm and proof are given in \Cref{sec:ReachDiam}. 

\begin{table}[ht]
  \centering
  \begin{threeparttable}
  \stepcounter{footnote}
  \centering
  \begingroup
  \setlength{\tabcolsep}{4pt}
  \renewcommand{\arraystretch}{1.05}
  \providecommand{\RDours}[1]{{\color{red}\bfseries\boldmath #1}}
  \newcommand{\RDcompactcell}[3]{%
    \begin{minipage}[c]{\linewidth}
      \vspace{1mm}
      \begin{minipage}[t][#1][t]{\linewidth}
        {\raggedleft\footnotesize upper\par}
        \vspace{1mm}
        \raggedright\normalsize #2
      \end{minipage}
      \par\hrule height 0.3pt\relax
      \vspace{1mm}
      \begin{minipage}[t][#1][t]{\linewidth}
        {\raggedleft\footnotesize lower\par}
        \vspace{1mm}
        \raggedright\normalsize #3
      \end{minipage}
    \end{minipage}%
  }
  \begin{tabular}{|m{0.17\linewidth}|m{0.23\linewidth}|m{0.45\linewidth}|}
    \hline
    \rule{0pt}{6mm}
    & \centering\arraybackslash\textbf{Undirected}
    & \centering\arraybackslash\textbf{Directed} \\[1mm]
    \hline
    \textbf{Unweighted}
    & \RDcompactcell{18mm}
        {$2$\quad\cite{dalirrooyfard2021diameter}}
        {$3/2-\varepsilon$\quad\cite{roditty2013diameter} (OV)}
    & \RDcompactcell{18mm}
        {prior: $n^{1/2+o(1)}$\quad\cite{bals2026diameter}\par\smallskip
         \RDours{ours: $n^{1/3+o(1)}$}}
        {prior: $3/2-\varepsilon$\quad\cite{bals2026diameter}\ (OV)\par\smallskip
         \RDours{ours: $n^{2/15-\varepsilon}$}\quad(HC)} \\
    \hline
    \textbf{Weighted}
    & \RDcompactcell{11mm}
        {$2$\quad\cite{dalirrooyfard2021diameter}}
        {$3/2-\varepsilon$\quad\cite{roditty2013diameter} (OV)}
    & \RDcompactcell{11mm}
        {---}
        {$n^c\ (\text{all }c>0)$\quad\cite{bals2026diameter}\ (SV)\tnote{\thefootnote}} \\
    \hline
  \end{tabular}
  \endgroup
  \caption{\textbf{Reachability diameter approximation in near-linear time.}
            Upper bounds correspond to $\widetilde O(m+n)$-time algorithms,
            lower bounds mean an approximation factor is excluded under the listed hypothesis, for fixed $\varepsilon>0$.
            For randomized HC the bound holds for algorithms with at least 2/3 success rate.
          }
  \label{tab:diameter-landscape}
  \begin{tablenotes}[flushleft]
  \footnotesize
  \item[\thefootnote] This near-linear-time consequence additionally
assumes $\omega>2$. Bals et al. rule out polynomial-factor
approximation in $O(n^{\omega-\delta})$ time for every fixed
$\delta>0$ under SV. Since $m\le n^2$, a
$\widetilde O(m+n)$-time algorithm would contradict this bound.
  \end{tablenotes}
  \end{threeparttable}
\end{table}

\paragraph{Polynomial approximation loss.}
Beyond breaking the $n^{1/2}$ barrier, the broader question of
Bals et al.~\cite{bals2026diameter} leaves open whether a constant-factor
approximation is possible in near-linear time.
They suggest approaching this problem from the other direction, 
proving a stronger lower bound,
a route which is especially compelling when comparing the approximation landscape for weighted and unweighted digraphs \Cref{tab:diameter-landscape}.
Recall that the reachability diameter measures the maximum finite distance between vertices of a digraph,
which coincides in the undirected setting with the maximum classic diameter of a connected component. 
In the undirected setting: applying well-known folklore algorithms to each connected component already gives a 2-approximation in near-linear time to the reachability diameter in both weighted and unweighted graphs.
Surprisingly, the picture is dramatically different in the directed setting: in the weighted regime, assuming $\omega > 2$ and the simplicial vertex hypothesis (SV), Bals et al. \cite{bals2026diameter} rule out \emph{every polynomial approximation factor} in near-linear time.
Yet switching to the unweighted regime changes the picture again.
Bals et al. obtain a polynomial-factor approximation in near-linear time, 
while their conditional lower bounds exclude only approximation factors below certain constants.
This motivates the following question:
\begin{problem}
    Does reachability diameter in unweighted digraphs admit a
    constant-factor, or more generally a subpolynomial-factor,
    approximation in near-linear time?
\end{problem}
Our result closes this question under the 3-uniform Hyperclique hypothesis (HC).
We show that polynomial approximation loss is unavoidable, 
although the optimal polynomial exponent remains open.

\begin{restatable}{theorem}{ReachDiamLowerBound}
\label{thm:lb-main}
Assume the deterministic $3$-uniform Hyperclique hypothesis.
For every fixed $\varepsilon\in(0,2/15)$, there is no deterministic
$\widetilde O(n+m)$-time algorithm that, on every unweighted
digraph $G$ with $n$ vertices, $m$ edges, and reachability diameter $D$,
returns an estimate satisfying
\[
    \frac{D}{n^{2/15-\varepsilon}}
    \leq \widehat D \leq D.
\]
Under the randomized version of the hypothesis, there is no
randomized $\widetilde O(n+m)$-time algorithm that, on every input
graph, returns an estimate satisfying this guarantee with
probability at least $2/3$.
\end{restatable}

One particular difficulty raised by Bals et al. \cite{bals2026diameter} is the ``triangle-inequality barrier" \cite{abboud2023diameter,karthik2020closest,rubinstein2018nearest}.
To establish a large approximation lower bound, a reduction must produce a large gap in diameter: one case must contain a pair of vertices far apart, while the other must keep every reachable pair close. The ``triangle-inequality barrier" obstructs the natural approach of creating a long path and selectively shortcutting its endpoints: intermediate vertices may remain far apart even after the shortcut is added, leaving both cases with large diameter.
Our reduction encodes a hyperclique candidate by a constant-direction trajectory in a layered DAG,
and adds short channels that compress trajectories encoding candidates that violate a hyperclique constraint.
When no hyperclique exists,
every reachable pair can be connected through a small number of the compressed trajectory segments.
An \EMPH{energy coordinate} preserves a long distance when a hyperclique does exist,
preventing any detours.
Finally, a Chinese remainder encoding keeps these channels small enough. Balancing the resulting trade-off between graph size and the diameter gap yields the exponent $2/15-\varepsilon$. 
The construction and proof are given in \Cref{sec:lower-bound}. 

\subsection{AI Disclosure}
ChatGPT 6 and 6.1 models contributed the majority of the
results,
and assisted with drafting and revision. 
The authors rewrote
all model-generated prose; the final wording is entirely their own.

%% file: Overview.tex
\section{Technical Overview}

\subsection{Preliminaries}
Let $G = (V, E)$ be the input unweighted digraph, with $n=|V|$ vertices and $m=|E|$ edges. 
Our algorithm proceeds by recursive partitioning;
to differentiate the base input graph from the input to an arbitrary recursive subproblem we write $H = (V(H), E(H))$ for the latter, 
where $H$ is a subgraph of $G$.

For $u, v \in V$, the function $d(u,v)$ is the shortest-path distance in $G$, 
even when $u$ and $v$ belong to a recursive subproblem.
In particular, $d(u, v) = \infty$ when there is no directed $u$-to-$v$ path in $G$, and $d(u, u) = 0$.
For a dipath $P$, we denote by $P[V(H)]$ the restriction of $P$ to $V(H)$. 
When $P = (u_{\text{start}}, \dots, u_{\text{end}})$ is a shortest dipath in $G$,
we use $d(P)$ to refer to the endpoint to endpoint distance $d(u_{\text{start}}, u_{\text{end}})$.

In any recursive instance with input $H$ and $u, v \in V(H)$, 
we write $u \leadsto_H v$ when there exists a directed $u$-to-$v$ path through $H$,
including dipaths consisting of a single vertex, 
so $u \leadsto_H u$.
When the relevant graph is clear from context, we simply say $u \leadsto v$.
The out-relevance $R_H^+(u) = \{v \in V(H) \mid u \leadsto v\}$ of $u$, also called the descendants of $u$ in $H$, is the subset of $V(H)$ which is reachable from $u$,
or equivalently, the set of vertices discovered during BFS initiated at $u$ on $H$.
Symmetrically, the in-relevance (ancestors) of $u$ in $H$, $R_H^-(u) = \{v \in V(H) \mid v \leadsto u\}$, is the set of vertices discovered during \emph{reverse} BFS: 
BFS with all edge directions reversed.
The union $R_H(u) = R_H^+(u) \cup R_H^-(u)$ is called the \emph{relevance}.
The size, or cardinality of these sets is denoted: $|R_H^+(u)|, |R_H^-(u)|, |R_H(u)|$.
For a dipath $P = (u_1, \dots)$ lying in $H$, the relevance is $R_H(P) = \cup_i R_H(u_i)$,
and similarly we have: $R_H^-(P) = \cup_i R_H^-(u_i)$, and $R_H^+(P) = \cup_i R_H^+(u_i)$.
When $V^\prime \subseteq V(H)$ is a set of vertices, $H[V^\prime]$ is the subgraph of $H$ induced by $V^\prime$. 

\subsection{Lower Bound}\label{sec:overview-lower-bound}

\paragraph{Polynomial approximation loss.}
The existing lower bounds for unweighted reachability diameter, based on
Boolean matrix multiplication and Orthogonal Vectors, establish
constant-factor approximation barriers~\cite{bals2026diameter}.
These results leave open whether a larger constant factor, or even a
polylogarithmic factor, can be achieved in near-linear time. In comparison,
the approximation guarantees of known near-linear-time algorithms,
including ours, still grow polynomially with $n$. \textit{This leaves a basic
question: is some polynomial approximation loss necessary, or can further
algorithmic improvements eliminate it?} We seek a conditional lower bound
showing that approximation within a fixed positive power of $n$ already
requires more than near-linear time. Such a bound would establish that
polynomial approximation loss is inherent under the hypothesis, while
leaving the optimal exponent open. Our lower bound result is built on the
following \EMPH{3-uniform Hyperclique hypothesis}.

A \EMPH{hypergraph} allows an edge to contain more than two vertices, which is called as a \EMPH{hyperedge}. In a 3-uniform hypergraph, every hyperedge consists of exactly three vertices. We consider a $K$-partite hypergraph whose vertex set is partitioned into $K$ disjoint parts \EMPH{$V_1,\ldots,V_K$}, each containing $N$ vertices. Every hyperedge contains vertices from three distinct parts. A \EMPH{$K$-hyperclique} is a set of $K$ vertices, one from each part, such that every three vertices in the set form a hyperedge.

\begin{hypothesis}[3-uniform Hyperclique Hypothesis~\cite{kunnemann2022klee}]
\label{hyp:overview-hyperclique}
For every fixed integer $K\geq4$ and constant $\eta>0$, there is no
deterministic algorithm deciding whether a 3-uniform $K$-partite
hypergraph with $N$ vertices in each part has a $K$-hyperclique in
$O(N^{K-\eta})$ time. The randomized version asserts the same for
randomized algorithms with error probability at most $1/3$ on every input.
\end{hypothesis}

Label the vertices in each part by \EMPH{$[N]:=\{0,\ldots,N-1\}$}, and let
\EMPH{$X_r\in[N]$} specify the vertex selected from $V_r$. Consider any three vertices $v_1,v_2,v_3$ from distinct parts $V_{r_1},V_{r_2},V_{r_3}$ that do not form a hyperedge. Since a $K$-hyperclique cannot contain all three vertices, we impose the constraint $\neg(X_{r_1}=v_1\wedge X_{r_2}=v_2\wedge X_{r_3}=v_3)$. Let \EMPH{$\mathcal C$} be the collection of all constraints obtained in this way. Then $|\mathcal C|\leq\binom K3N^3$. Trivially, we have
$$
\begin{aligned}
 \text{a hyperclique exists}
   &\iff \text{some assignment violates no constraint},\\
 \text{no hyperclique exists}
   &\iff \text{every assignment violates at least one constraint}.
\end{aligned}
$$

\paragraph{Choosing the hypothesis.}
Our choice of the Hyperclique hypothesis is motivated by a connection
between reachability diameter and geometric coverage. A small reachability
diameter means that short paths connect every reachable pair, while a
large reachability diameter witnesses a failure of this property. This
resembles geometric coverage, where the question is whether a collection
of objects covers an entire region or leaves an uncovered point.
Hyperclique-based lower bounds for Klee's measure problem exploit precisely
this distinction: complete coverage corresponds to the absence of a
hyperclique~\cite{kunnemann2022klee}.

In this analogy, an uncovered point corresponds to a reachable pair whose
distance remains large. The related shortcutting problem asks whether
adding a small set of reachability-preserving edges can make every
reachable pair close. Shortcut lower bounds show that making all reachable
pairs close can face a polynomial obstruction~\cite{huangpettie2018shortcuts}.
We ask whether detecting a remaining long distance also faces a polynomial
approximation barrier in near-linear time. The Hyperclique hypothesis
provides a natural starting point for connecting the computational hardness
of global coverage with the structural obstacles identified by shortcut
lower bounds.

\paragraph{Constructing the graph.}
\textit{Our goal is to construct a digraph substantially faster than
enumerating the $N^K$ candidates, with a large reachability diameter if and
only if a hyperclique exists.} Our digraph is built on geometric
constructions used in shortcut and distance-preserver lower
bounds~\cite{hesse2003shortcuts,coppersmithelkin2006preservers,huangpettie2018shortcuts,bodwinhoppenworth2024hopsets}.
We assign integer coordinates to vertices and arrange them in layers,
with all edges directed from earlier layers to later layers. \textit{Thus, the
graph is a DAG and \Cref{thm:lb-main} also applies to DAGs.} Each trajectory
follows a fixed direction across the layers and encodes a candidate
assignment to $X_1, \dots, X_K$. We connect its vertices in consecutive
layers by constant-length dipaths. If the encoded assignment violates a
constraint in $\mathcal{C}$, we also add shortcuts that cross many layers
using only a constant number of edges to force short distances.

\paragraph{Protecting a satisfying assignment.}
The main difficulty is to ensure that shortcuts for failing assignments do
not also shorten distances associated with a satisfying assignment.
Earlier shortcut and distance-preserver lower bounds choose movement
vectors so that none is a weighted average of the
others~\cite{hesse2003shortcuts,coppersmithelkin2006preservers,huangpettie2018shortcuts}.
We adapt this convexity argument using an energy coordinate. More
specifically, each move between consecutive layers increases this
coordinate by the squared Euclidean norm of its direction, and each
shortcut contributes the total increase of the moves it replaces.

Consider different sequences of moves that cross the same number of
layers and produce the same total change in the spatial coordinates.
Moving in a constant direction gives the smallest total energy increase.
Any sequence that changes direction requires a strictly larger energy
increase. This comparison is closely related to the use of squared
movement costs to exclude alternative shortest paths in Bodwin and
Hoppenworth~\cite[Section~4.4, Proposition~3 and Lemma~2]{bodwinhoppenworth2024hopsets}.
We set the energy difference between vertices on a trajectory to this
minimum. Any dipath connecting them must therefore preserve the
trajectory's direction and cannot use shortcuts for a different assignment.
Consequently, a satisfying assignment retains a distance proportional to
the number of layers.

\subsection{Upper Bound}

Our algorithm builds on the recursive partitioning framework of Jambulapati, Liu, and Sidford \cite{liu2019reachability}. 
We call the following recursive sampling, searching, labelling, and grouping procedure \emph{JLS partitioning}.

\paragraph{JLS partitioning.} 
Let $H$ be the input to an arbitrary instance at level $r$ of the recursion, 
where the root instance $G$ is at level $r=0$.

\begin{itemize}
    \item \textbf{Sampling:} Sample pivots $S$ from $V(H)$ uniformly at random with probability $p_r$. 
    The sampling probability grows by $\Theta(\log n)$ per recursion level, 
    so after $O(\log n)$ levels $p_r = 1$ and every remaining vertex becomes a pivot.
    \item \textbf{Search:} Run BFS and reverse BFS from each $s \in S$. 
    Remove all vertices found in both search directions initiated at a pivot.
    \item \textbf{Labelling:} Each pivot $s$ assigns every $v \in V(H)$ a \emph{reachability signature} 
    \[
        \sigma_s(v) := \bigl(\mathbf{1}[v\leadsto s], \mathbf{1}[s\leadsto v] \bigr).
    \]
    where $\mathbf{1}[\cdot]$ denotes the indicator function. 
    \item \textbf{Grouping:} Group the remaining vertices into sets $V_1, \dots$ according to their reachability signatures. 
    The induced subgraphs $H[V_1], \dots$ are the inputs to the recursive children of $H$. 
\end{itemize}

We now make precise what it means to \emph{resolve} a dipath fragment. 
To complete the construction of Jambulapati et al. \cite{liu2019reachability},
for each pair consisting of a recursive input $H$ and an associated pivot~$s$ add the following edges to the shortcut set:
\[
    \text{i) } (u,s) \text{ for every } u \in R_H^-(s),
    \qquad
    \text{ii) } (s,v) \text{ for every } v \in R_H^+(s).
\]
These edges preserve reachability by definition.
Now fix some instance with input $H$ and pivot samples $S \subseteq V(H)$. 
Let $P$ be any $u$-to-$v$ dipath in $H$, $u, v \in V(H)$.
A pivot $s$ is called a \emph{bridge} of $P$ when $u \leadsto s$ and $s \leadsto v$.
If such a pivot exists, 
then the shortcut edges $(u, s)$ and $(s, v)$ reduce the shortest $u$-to-$v$ distance to at most two hops in the augmented graph.
We therefore regard $P$ as resolved for the shortcut construction.
In particular, 
vertices removed during the search phase are mutually reachable with a sampled pivot, which bridges any path fragments consisting of those vertices.
After $O(\log n)$ levels,
the sampling probability satisfies $p_r = 1$ and every remaining vertex is sampled as a pivot,
including every vertex lying on every remaining path, thereby resolving them.

\paragraph{From shortcuts to detecting distances.} 
Bals et al. \cite{bals2026diameter} connect resolving a shortest-path fragment with detecting a large distance in the original graph through the following structural guarantees of JLS partitioning.
\begin{lemma}[Adapted from Lemma~22 and Corollary~23 of Bals et al. \cite{bals2026diameter}]\label{lem:contiguous-subpath}
    Let $G$ be an unweighted digraph, 
    and let $H$ be the input to any recursive instance obtained from $G$ by repeated JLS partitioning.
    Then:
    \begin{itemize}
        \item For every directed path $P$ in $G$,
        the restriction $P[V(H)]$ is either empty or consists of a contiguous subpath of $P$.
        \item For every $u,v \in V(H)$, the shortest-path distance from $u$ to $v$ in $H$ equals $d(u, v)$, 
        their shortest-path distance in $G$.
    \end{itemize}
\end{lemma}
\begin{proof} 
    The base case $H=G$ is immediate.  
    Now suppose that an arbitrary instance with input $H$ satisfies both given statements. 
    We proceed by induction on the recursive subproblems $H[V_1], ...$.
    By way of contradiction, suppose that $u, v, w \in V(H)$, all different, such that there is a $u$-to-$v$ directed path in $H$ containing $w$, 
    but after partitioning $w$ end up in a different child than $u, v$: $u, v \in V_i$ and $w \notin V_i$.
    The case where $w$ is dropped after the searching phase is immediate,
    if there exists some pivot such that $w$ is found in both of its search directions then $u$ and $v$ are also found in both searches, 
    and dropped.
    Since $u$ and $v$ belong to the vertices of the same child they have identical reachability signatures.
    Meaning that, for every sampled pivot $s\in S$, 
    we have $s\leadsto u\Rightarrow s\leadsto w\Rightarrow s\leadsto v$ and $v\leadsto s\Rightarrow w\leadsto s\Rightarrow u\leadsto s$.
    These implications force $u$ and $v$ to have exactly the same labels as $w$, a contradiction.
    Thus, 
    for every ordered pair of endpoints that are grouped into the same recursive child, 
    every vertex on every dipath connecting them in $H$ is grouped into the same child.
    By induction, 
    every directed path $P$ in $H$,
    the restriction $P[V_i]$ is either empty or consists of a contiguous subpath of $P$.
    Moreover, 
    for every $u,v \in V_i$, the shortest-path distance from $u$ to $v$ in $H[V_i]$ equals $d(u, v)$, 
    their shortest-path distance in $G$.
\end{proof}
\begin{observation}[Observation~21 and Corollary~23 of Bals et al. \cite{bals2026diameter}]\label{obs:bridge-metric}
    Let $G$ and $H$ be as in \Cref{lem:contiguous-subpath}.
    Then for $u, v \in V(H)$, 
    the shortest-path distance from $u$ to $v$ in $H$ equals their shortest-path distance in $G$.
    In other words, 
    if $d(u, v)$ is finite, 
    then $H$ contains some $u$-to-$v$ dipath $P$ of length $d(u, v)$.
    Fix some such $u, v$, 
    and suppose that we sample a bridge $s$ of $P$ in the recursive instance corresponding to input $H$, 
    so $u \leadsto s$ and $s \leadsto v$.
    By another application of \Cref{lem:contiguous-subpath},
    BFS and reverse BFS from $s$ in the search phase of JLS partitioning reveal $d(u, s)$ and $d(s, v)$.
    The triangle inequality gives: $d(u, v)/2 \leq \max(d(u, s), d(s, v)) \leq D$.
    Thus, resolving $P$ reveals a distance at least half its length,
    without overestimating the reachability diameter of $G$.
\end{observation}

This observation explains how Bals et al. acquire distance estimates from resolved fragments, 
inheriting the $n^{1/2+o(1)}$ bound of the underlying shortcut analysis.
Our approach seeks to obtain an estimate during the earlier stages of the JLS partitioning,
while fragments of diametral paths retain more of their length.
However, stopping earlier in the analysis requires a method to detect distances in unresolved fragments, 
for which the argument of \Cref{obs:bridge-metric} provides no guarantee.

\paragraph{Detecting distances in unresolved fragments.} 
For unresolved path fragments, 
our algorithm adapts a procedure of Cohen used for estimating the sizes of reachability sets (Algorithm 3.1 of \cite{cohen1997size}).
The general description of the algorithm is as follows: process the vertices of an input digraph in a uniformly random order, 
using reverse BFS to identify their \emph{least-rank descendants}. 
The idea is that vertices with larger out-relevance sets expect to see their first descendant earlier in the ordering,
so the minimum rank of its descendants in the ordering informs an estimate of the size of the out-relevance.

Cohen also establishes that the choice of least-descendant is distributed uniformly at random over the associated out-relevance set (Propositions 2.1-2.2 of \cite{cohen1997size}).
Our $\Call{Sweep}{}$ subroutine exploits this fact in the symmetric procedure, 
which identifies the least-rank \emph{ancestor}, 
to detect distances in unresolved fragments. 
\begin{algorithm}[H]
    \caption{$\textsc{Sweep}(H)$.}
    \label{alg:Sweep}
    \textbf{Input:} Recursive instance $H$.\par
    \textbf{Output:} An original distance in the base input graph to $\Call{ReachDiam}{}$.
    \begin{enumerate}
        \item Process the initially uncolored vertices $V(H)$ in a uniformly random order. When an uncolored vertex is encountered: start BFS with it as the root, then color and remove every vertex discovered by the search (in the working graph only).
        \item Report the largest distance observed during any search.
    \end{enumerate}
\end{algorithm}
\noindent As in Cohen's procedure,
each vertex is colored by the search rooted at its first ancestor encountered during processing,
its least-rank ancestor in $H$.

To see why $\Call{Sweep}{}$ works to detect distance in unresolved fragments:
let $H$ be the input to any recursive instance obtained from $G$ by repeated JLS partitioning,
such that $H$ contains a shortest dipath $P = (p_1, \dots, p_{k})$ with in-relevance $R^-_H(P)$. 
Suppose that we do \textbf{not} sample a bridge of $P$ in the current instance,
i.e. $P$ is unresolved in this exercise.
First, we naturally extend the definition of least-rank ancestor to dipaths $P$ 
as the first ancestor of any of its vertices encountered during processing.
In particular, Propositions 2.1-2.2 of \cite{cohen1997size} still apply:
the least-rank ancestor $a$ of $P$ is distributed uniformly at random over $R^-_H(P)$.
Additionally, $a$ and every vertex of $P$ are uncolored at the moment when $a$ is encountered in the ordering,
otherwise we must have already processed an ancestor of $P$,
a contradiction with $a$'s minimality.
Now consider the event that $a$ is one of the vertices $p_1, \dots, p_{\lfloor k/2\rfloor}$ lying on the first half of $P$,
called \emph{witnesses}.
Then BFS rooted at $a$ reveals the length of the $a$-to-$p_k$ subpath of $P$, 
which is at least half the length of $P$.
By \Cref{lem:contiguous-subpath}, 
the length of $P$ in $H$ is equal to the shortest-path distance $d(p_1, p_k)$ in $G$,
thus $\Call{Sweep}{}$ reports at least half the length of the unresolved path fragment.
Moreover,
$\Call{Sweep}{}$ does not overestimate.
After each search the colored set is closed under reachability in $H$,
and therefore all paths between uncolored vertices are themselves uncolored,
meaning that BFS distances reported by $\Call{Sweep}{}$ equal their distances in $H$, hence $G$.
Then, 
conditioned on the event described above, 
$\Call{Sweep}{}$ reports an original distance in the input at least half the length of the unresolved path fragment.

Moreover, 
the choice of $a$ is uniformly random over $R^-_H(P)$,
so the probability that $a$ lies on the first half of $P$ is exactly the ratio of the number of witnesses $p_1, \dots, p_{\lfloor k/2 \rfloor}$, 
to the total size of $P$'s in-relevance in $H$.

\paragraph{Controlling the ratio of witnesses.} 
The analysis of JLS partitioning by Ashvinkumar et al. \cite{ashvinkumar2026parallel} provides the missing ingredient:
for every path, a bound on the number of recursive inputs containing a subpath with large relevance,
allowing us to establish a trade-off between the amount of fragmentation and the success probability of $\Call{Sweep}{}$. 
Let $G$ be the input to JLS partitioning, and $P$ a fixed shortest dipath in $G$.
By \Cref{lem:contiguous-subpath}, every recursive instance whose input subgraph contains vertices of $P$ contains a contiguous subpath of $P$.
Now consider the \emph{subproblem tree} of $P$, 
whose nodes are those recursive instances sorted by level,
and whose edges link parent to child.
Instances in the subproblem tree in which a fragment is resolved are treated as leaves: 
by the argument of \Cref{obs:bridge-metric}, resolving a shortest-path fragment reveals at least half its length without overestimating,
so in either case we stop tracking its further subdivisions in the analysis.
The following lemma
\footnote{Ashvinkumar et al. state these results for DAGs, but their arguments also apply to general digraphs. The underlying mechanisms, Lemmas IV.1 and IV.4 of \cite{liu2019reachability}, are already stated for general digraphs.}
of Ashvinkumar et al. \cite{ashvinkumar2026parallel}
bounds the expected number of nodes in the subproblem tree where the associated fragment of $P$ has large relevance:
\begin{lemma}[Adapted from Lemma~4.5 and Observation~4.6 in~\cite{ashvinkumar2026parallel}]\label{lem:large-node-bound-fanout-bound}
    Fix a shortest path $P$ in the input to JLS partitioning, and a threshold $x > 0$.
    Let $X$ be the count of nodes in the subproblem tree of $P$
    whose associated fragment $P^\prime = P[V(H)]$ satisfies $|R_H(P^\prime)| \geq x$,
    where $H$ is the corresponding recursive input.
    Then:
    \[
    \mathbb{E}[X] \leq \sqrt{\frac{n^{1+O(1/\log{\log n)}}}{x}}.
    \]
    Moreover, 
    with probability at least $1-O(n^{-10})$,
    every node in the subproblem tree of $P$ has $O(\log^2{n})$ children.
\end{lemma}

Taken together,
these bounds control the expected fragmentation of a diametral path, 
stopping the analysis of each subproblem when it's fragment is resolved or when its relevance becomes sufficiently small.
We show that the resulting fragments collectively yield a constant probability of success of detecting a distance at least $D/n^{1/3+o(1)}$,
either through a sampled bridge or $\Call{Sweep}{}$,
achieving the desired sub-$n^{1/2}$ approximation factor.

%% file: LowerBound.tex
\section{A Polynomial Lower Bound for Reachability Diameter}\label{sec:lower-bound}

We refer to the technical overview in \Cref{sec:overview-lower-bound}
for the 3-uniform Hyperclique hypothesis and the formulation of
Hyperclique using variables $X_1,\ldots,X_K$ and constraints
$\mathcal C$.

\subsection{Main ideas and graph construction framework}\label{sec:lb-trajectories}

\paragraph{Encoding candidate assignments in the graph.}

We now explain how to encode the candidate assignments for $X_1, \dots, X_K$ in our layered graph. At a high level, each \EMPH{trajectory} in the graph encodes an assignment to the variables $X_1,\ldots,X_K$.

We begin by outlining the graph construction. Consider a vertex represented by the coordinates $(\tau,x_1,x_2,\ldots)$, where \EMPH{$\tau$} denotes its layer and \EMPH{$\mathbf x=(x_1,x_2,\ldots)$} gives its coordinates within that layer. Now, consider two vertices $(\tau_1,\mathbf x)$ and $(\tau_2,\mathbf y)$ with $\tau_1<\tau_2$. If $(y_i-x_i)/(\tau_2-\tau_1)$ is an integer for every $i$, we can obtain the second endpoint from the first by advancing one layer at a time and changing the $i$-th coordinate by exactly $(y_i-x_i)/(\tau_2-\tau_1)$ at each step. Thus, the two vertices determine a constant direction across these layers. We call such a pair of vertices a \EMPH{valid pair}.

For a valid pair $((\tau_1,\mathbf x),(\tau_2,\mathbf y))$, the \EMPH{direction} is \EMPH{$\mathbf a=(\mathbf y-\mathbf x)/(\tau_2-\tau_1)$}. Let \EMPH{$\mathbf b = \mathbf x-\tau_1\mathbf a=\mathbf y-\tau_2\mathbf a$} denote the \EMPH{intercept} of the pair. By fixing $\mathbf{a}$ and $\mathbf b$, each valid pair determines dipaths starting from $(0,\mathbf b)$ that follows a constant direction. We call such dipaths follow the same \EMPH{trajectory}. 

Thus, the amount of information a trajectory can encode is determined by the possible choices of $\mathbf b$ and $\mathbf a$. Hence, all the information carried by a trajectory is contained in $\mathbf b$ and $\mathbf a$, so encoding an assignment in a trajectory amounts to encoding it in these two vectors. In the meanwhile, the number of possible choices of $\mathbf b$ and $\mathbf a$ determines how many distinct trajectories we can use. By assigning different blocks to different variables, we can distribute the values of $X_1,\ldots,X_K$ across the starting coordinates $\mathbf b$ and the direction components $\mathbf a$. Thus, we ensure that every candidate assignment to $X_1, \dots, X_r$ is represented by a trajectory whose vertices all belong to the graph.

\paragraph{Compressing a failing trajectory.}

Recall that our goal is to construct a digraph whose reachability diameter is large if and only if there exists an assignment to $X_1,\ldots,X_K$ satisfying every constraint in $\mathcal C$.

Consider any valid pair in the digraph. The two vertices uniquely determine the intercept $\mathbf b$ and direction $\mathbf a$ of the corresponding trajectory. Starting from $(0,\mathbf b)$ and moving in direction $\mathbf a$ from one layer to the next, we obtain a unique vertex $(\tau,\mathbf b+\tau\mathbf a)$ in each layer $\tau$. We call these vertices as \EMPH{vertices on the trajectory}. In our construction, we ensure that the vertices on each trajectory in any two consecutive layers are connected by a constant-length dipath from the earlier layer to the later layer. Note that any pair of vertices on the trajectory is a valid pair by definition. 

If the trajectory determined by $\mathbf b$ and $\mathbf a$ encodes an assignment that violates a constraint in $\mathcal C$, we call it a \EMPH{failing trajectory}. We want every pair of vertices on a failing trajectory to be connected by a short path from the earlier layer to the later layer. Thus, a failing trajectory should not force a large diameter. A simple solution is to add a shortcut from every earlier vertex to every later vertex on the trajectory, but enumerating all pairs would be too expensive.

To reduce the number of pairs we need to connect directly, we choose
a collection of intervals of layers. Let \EMPH{$L\geq2$} be a power of two, and denote the total number of layers. For each length $\ell\in\{1,2,4,\ldots,L\}$, we choose the intervals $I=[j\ell,(j+1)\ell]\subseteq[0,L]$, where $j$ is a nonnegative integer. For example, when $\ell=2$, we choose $[0,2],[2,4],\ldots, [L-2,L]$. Each interval has length equal to a power of two and starts at a
multiple of its length, which is called as a \EMPH{dyadic interval}.

For a failing trajectory, we will construct a \EMPH{constant-length shortcut} between any pair of vertices on the trajectory in each dyadic interval. We also ensure that each shortcut has the same direction as the trajectory determined by the two vertices it connects.

Fix any two vertices on the trajectory, the interval between their layers can be partitioned into at most $2\log L$ dyadic intervals. To see this, consider any interval $[s,t]\subseteq[0,L]$. Start with the unit intervals between $s$ and $t$, and repeatedly merge two adjacent intervals whenever their union is a dyadic interval, until no further merges are possible. Given the construction of the dyadic intervals, at most of two intervals of each length remain. Thus, the resulting partition contains only $O(\log L)$ dyadic intervals. 

In conclusion, we only need to construct constant-length shortcuts 
for the $2L-1$ aligned dyadic intervals in $[0,L]$ to obtain short
paths between every pair of vertices on a failing trajectory.

\paragraph{Protecting a satisfying trajectory.}
Consider a trajectory determined by the intercept $\mathbf b$ and direction $\mathbf a$. Suppose its encoded assignment satisfies every constraint in $\mathcal C$, and such trajectory is a \EMPH{satisfying trajectory}. We want the distance from its vertex in layer $0$ to its vertex in layer $L$ to remain large such that we force the existence of a large reachability diameter in this case. 

However, our current construction cannot yet guarantee that the distance between two vertices remains large. A valid pair determines a fixed intercept and direction, but a dipath between them need not keep that direction throughout. Since different trajectories may share vertices, a dipath may change direction along the way and still reach the prescribed final vertex. If a segment of the dipath connects vertices on a failing trajectory, that segment may use the shortcuts constructed above. Thus, even though the original pair determines a satisfying trajectory, a dipath between the two vertices may still use shortcuts for failing assignments.

To prevent such a dipath, we need to ensure that a dipath connecting a valid pair on the corresponding trajectory cannot switch to a different direction.

Recall that moving in direction $\mathbf a$ from one layer to the next changes $\mathbf x$ by $\mathbf a$. We add a coordinate \EMPH{$z$} to each vertex, called the \EMPH{energy coordinate}. For every move in direction $\mathbf a$ between consecutive layers, we require $z$ to increase by $\|\mathbf a\|^2$. Just as $\mathbf b$ gives the coordinates $\mathbf x$ in layer $0$, let \EMPH{$e$} denote the value of $z$ in layer $0$. Starting from $(0,\mathbf b,e)$ and applying these changes from one layer to the next, we obtain the vertex on the trajectory in layer $\tau$ with coordinates
\begin{equation}\label{eq:lb-trajectory}
   (\tau,\mathbf x(\tau),z(\tau))
       =(\tau,\mathbf b+\tau\mathbf a,
                    e+\tau\|\mathbf a\|^2).
\end{equation}
Thus, with the energy coordinate included, $\mathbf b$, $e$, and $\mathbf a$ determine a trajectory.

Recall that a constant-length shortcut between any two vertices on this trajectory also follows direction $\mathbf a$. If a constant-length shortcut is added between two vertices on this trajectory in layers $\tau_1<\tau_2$, we require the total increase in $z$ along the shortcut to be $(\tau_2-\tau_1)\|\mathbf a\|^2$. Thus, the energy coordinate changes from $e+\tau_1\|\mathbf a\|^2$ to $e+\tau_2\|\mathbf a\|^2$, exactly as it would when moving in direction $\mathbf a$ from one layer to the next.

We next explain why the energy coordinate prevents a dipath between two vertices on a trajectory from switching directions. 

The argument adapts the convexity proof in~\cite[Section~2.1, Lemma~2]{huangpettie2018shortcuts}. Fix two vertices on the trajectory determined by $\mathbf b$, $e$, and $\mathbf a$, and consider a dipath from the earlier vertex to the later vertex. Suppose for contradiction that a dipath between does not keep the same direction of $\mathbf a$ throughout.Thus, we can divide the dipath into segments, each of which follows a fixed direction throughout. A segment may contain any number of moves between consecutive layers or shortcuts across dyadic intervals, provided that they all have the same direction. Let $\mathbf a_1,\ldots,\mathbf a_r$ be the directions of these segments. For segment $j$, let $\ell_j>0$ be the difference between the layer indices of its first and last vertices.

Thus, segment $j$ changes $\mathbf x$ by $\ell_j\mathbf a_j$ and $z$ by
$\ell_j\|\mathbf a_j\|^2$. Let $T=\sum_j\ell_j$ be the difference
between the layers of the two fixed vertices. Since both vertices lie on the trajectory with direction $\mathbf a$, the total changes in $\mathbf x$ and $z$ must also be $T\mathbf a$ and $T\|\mathbf a\|^2$, respectively. Hence,
$$
 \sum_j\ell_j\mathbf a_j=T\mathbf a,
 \qquad
 \sum_j\ell_j\|\mathbf a_j\|^2=T\|\mathbf a\|^2.
$$
Consequently,
\begin{equation}\label{eq:lb-rigidity}
 \sum_j\ell_j\|\mathbf a_j-\mathbf a\|^2
 =T\|\mathbf a\|^2-2T\|\mathbf a\|^2+T\|\mathbf a\|^2=0.
\end{equation}
Every term in the sum is nonnegative, and every $\ell_j$ is positive. Hence, $\mathbf a_j=\mathbf a$ for every segment $j$. In conclusion, every dipath connecting the a valid pair must always follow the direction of the trajectory determined by that pair. 

Now, consider a valid pair with one vertex in layer $0$ and the other in layer $L$, and suppose the trajectory determined by this pair is a satisfying trajectory. No dipath between the two vertices can use a shortcut belonging to a different trajectory, since using such a shortcut would require changing direction.

Recall that after adding the energy coordinate, we need the intercept $\mathbf b$, the direction $\mathbf a$, and $e$ to specify a trajectory.  We also use $e$ to encode some part of the variables from $X_1, \dots, X_K$, just as how we use the $\mathbf b$ and $\mathbf a$. Otherwise, the energy coordinate would enlarge the graph without increasing the number of assignments we can encode, weakening the reduction.

\paragraph{Choosing the number of coordinates.}
Recall that we encode the values of $X_1,\ldots,X_K$ in the intercept $\mathbf b$, the direction $\mathbf a$, and $e$. We have not yet specified how many coordinates $\mathbf x$ has. Since $\mathbf b$ and $\mathbf a$ each have one component for every coordinate of $\mathbf x$, adding one coordinate gives us two more components in which to encode variable values. 

However, more coordinates also increase the size of
the graph and the cost of constructing it. Thus, we need to balance the number of assignments we can encode against the size of the graph and the time needed to construct it. We choose three coordinates, $\mathbf x=(x_1,x_2,x_3)$, together with the energy coordinate $z$, so that $\mathbf b$ and $\mathbf a$ each
have three components. Our analysis shows that using three coordinates gives the strongest lower bound on the approximation ratio for this construction. We justify this choice in \Cref{sec:lb-parameters}, after deriving these size bounds.

We summarize the construction described so far in \Cref{alg:lb-framework}. In the next subsection, we describe how to add edges to connect vertices in consecutive layers and how to implement the constant-length shortcuts.

\begin{algorithm}[H]
\caption{Graph Construction Framework}\label{alg:lb-framework}
\renewcommand{\algorithmicrequire}{\textbf{Input:}}
\begin{algorithmic}[1]
\setlength{\itemsep}{0.15em}
\Require Constraints $\mathcal C$ on $X_1,\ldots,X_K$ and $L\geq2$
\State Choose bounded ranges for $\mathbf x=(x_1,x_2,x_3)$ and the direction $\mathbf a$.
\State \parbox[t]{\linewidth}{\raggedright
  Create vertices $(\tau,\mathbf x,z)$ in layers $\tau=0,\ldots,L$,
  with coordinates restricted to the chosen ranges.}
\State Encode each assignment to $X_1,\ldots,X_K$ in $\mathbf b$, $\mathbf a$, and $e$.
\State \parbox[t]{\linewidth}{\raggedright
  Connect vertices on the same trajectory in consecutive layers:\\[0.35ex]
  \hspace*{\algorithmicindent}
  $(\tau,\mathbf x,z)\longrightarrow
  (\tau+1,\mathbf x+\mathbf a,z+\|\mathbf a\|^2)$.}
\For{each dyadic interval $I=[\tau,\tau+\ell]$ with $\ell\geq2$}
  \State \parbox[t]{\dimexpr\linewidth-\algorithmicindent\relax}{\raggedright
    For a failing trajectory, add a shortcut
    between vertices in layers $\tau$ and $\tau+\ell$:\\[0.35ex]
    \hspace*{\algorithmicindent}
    $(\tau,\mathbf x,z)\longrightarrow
    (\tau+\ell,\mathbf x+\ell\mathbf a,z+\ell\|\mathbf a\|^2)$.}
\EndFor
\end{algorithmic}
\end{algorithm}

\subsection{Connecting consecutive layers and adding shortcuts}\label{sec:lb-unoptimized}

Recall that we need to connect vertices on the same trajectory in
consecutive layers, and to add constant-length shortcuts across dyadic
intervals for failing trajectories. To realize this, We first choose the
direction components one at a time to connect consecutive layers. We
then use the same construction for shortcuts, adding checks to ensure
that a shortcut is available only when the encoded assignment violates
a constraint in $\mathcal C$.

First, we will specify the coordinate ranges.

\paragraph{Choosing the coordinate ranges.}
Let \EMPH{$q$} be the number of choices for each direction component, so that
$0\leq a_i<q$ for each $i\in\{1,2,3\}$. Recall that advancing one layer
increases $x_i$ by $a_i$ and $z$ by $\|\mathbf a\|^2$. Over $L$ layers,
each $x_i$ increases by less than $Lq$, and $z$ increases by
less than $3Lq^2$. Thus, if we choose $0\leq b_i<Lq$ and
$0\leq e<Lq^2$, every vertex on the trajectory from layer $0$ to
layer $L$ satisfies $0\leq x_i<2Lq$ and $0\leq z<4Lq^2$.

In each layer $\tau=0,\ldots,L$, we create all vertices $(\tau,x_1,x_2,x_3,z)$ with integer coordinates satisfying $0\leq x_i<2Lq$ for $i=1,2,3$ and $0\leq z<4Lq^2$. We call them \EMPH{base vertices} to distinguish them from the \EMPH{auxiliary vertices} introduced later. For
convenience, write \EMPH{$B=Lq^2$} and let $U$ denote the set of allowed coordinates $(\mathbf x,z)$: \EMPH{$U=[2Lq]^3\times[4B]$}. In particular, $|U|=32L^4q^5$.

\paragraph{Connecting consecutive layers.}
Consider a base vertex in layer $\tau$. There are $q^3$ choices of $\mathbf a$, so adding a separate edge for every direction would be too expensive. 

Instead, we introduce two \EMPH{auxiliary layers} between layers $\tau$ and $\tau+1$, each containing one vertex for every $(\mathbf x,z)\in U$. We add edges from layer $\tau$ to the first auxiliary layer, then to the second auxiliary layer, and finally to layer $\tau+1$. The $i$-th edge of each resulting three-edge dipath chooses $a_i$ and updates only $x_i$ and $z$. More specifically, the $i$-th edge updates $x_i$ to $x_i+a_i$ and $z$ to $z+a_i^2$. 

A three-edge dipath combines one choice of $a_1$, one of $a_2$, and one of $a_3$. These combinations represent all $q^3$ directions without requiring us to enumerate them explicitly when constructing the edges. Thus, each vertex needs only $q$ choices for its outgoing edges. Moreover, the three-edge dipath gives the required total changes in $\mathbf x$ and $z$. Conversely, consider two base vertices on the same trajectory in consecutive layers, with $0\leq a_i<q$. Choosing their direction components in the three stages gives a dipath between them. Every intermediate coordinate lies between its values at the two base vertices, so all intermediate vertices belong to the corresponding copies of $U$.

This construction is inspired by the alternation product of Huang and Pettie~\cite[Section~2.2]{huangpettie2018shortcuts}, which builds on Hesse's construction~\cite{hesse2003shortcuts}.

\paragraph{Add constant-length shortcuts.}
For a dyadic interval $I=[\tau,\tau+\ell]$ with $\ell\geq2$, we add a constant-length shortcut in a similar way with changes $\ell\mathbf a$ and $\ell\|\mathbf a\|^2$ in $\mathbf x$ and $z$. However, we want a shortcut only when the trajectory determined by the two base vertices is failing.  

The difficulty is that we construct the three edges of a shortcut by choosing $a_1$, $a_2$, and $a_3$ separately. When constructing the edge that chooses $a_i$, we can recover $b_i=x_i-\tau a_i$, and decode the variables stored in $a_i$ and $b_i$. How can these variables collectively determine whether the complete assignment violates a constraint in $\mathcal C$ without exhaustive enumeration? 

Note that an assignment is failing if it violates at least one constraint in $\mathcal C$. Thus, we can test the constraints separately. For each $C \in\mathcal C$, we create a \EMPH{channel} with its own two auxiliary layers. Our goal is to allow a complete three-edge dipath through this channel exactly when the encoded assignment violates $C$. Thus, a failing assignment only needs to allow a dipath through one of these channels.

To violate $C$, all three variables involved in $C$ must equal their corresponding forbidden values. When constructing an edge that chooses $a_i$, we check every variable of $C$ stored in $a_i$ or $b_i$, and add the edge only if all their decoded values match the forbidden values specified by $C$. If neither $a_i$ nor $b_i$ stores a variable of $C$, we keep the edge without imposing any condition from $C$. Since each channel has its own auxiliary vertices, a dipath through the channel must pass the checks for the same constraint on all three edges. These checks ensure that every variable of $C$ stored in $\mathbf a$ or $\mathbf b$ has its forbidden value. 

To ensure that a complete dipath is available only when $C$ is violated, we must also check any variables of $C$ stored in $e$. We explain how to perform this remaining check next.

Unlike $b_i$, the value of $e$ depends on all three direction components  since $e=z-\tau\|\mathbf a\|^2$. We cannot recover $e$ from the choice of a single component of the direction. Instead, we guess the energy coordinate of the base vertex in layer $\tau+\ell$. Together with the energy coordinate of the base vertex in layer $\tau$, this guess determines $e$. On the edge leaving the base vertex in layer $\tau$, we use the resulting value of $e$ to check the variables of $C$ stored in $e$. On the edge entering the base vertex in layer $\tau+\ell$, we require the energy coordinate reached by that edge to equal the guess.

To see why this works, write $\tau=j\ell$, where $j$ is an non-negative integer. Let $z_0$ and $z_1$ denote the energy coordinates of the two base vertices in layers $\tau$ and $\tau+\ell$, respectively. A complete three-edge dipath satisfies $z_1-z_0=\ell\|\mathbf a\|^2$, so
\begin{equation}\label{eq:lb-energy-identity}
 e=z_0-\tau\|\mathbf a\|^2=(j+1)z_0-jz_1.
\end{equation}

We now explain how to modify the graph construction accordingly. For each $C\in\mathcal C$, create a separate channel for every integer guess $0\leq z <4B$ of $z_1$. On the edge leaving layer $\tau$, use $(j+1)z_0-jz$ as the value of $e$ and require all variables of $C$ stored there to equal their forbidden values. On the edge entering layer $\tau+\ell$, we require the energy coordinate $z_1$ of the base vertex reached by the shortcut to equal the guessed value $z$. Since channels for different guesses have separate auxiliary vertices, the guess verified at layer $\tau+\ell$ must be the same one used to compute $e$ on the edge leaving layer $\tau$. Thus, a complete three-edge dipath through the channel guarantees that the variables of $C$ stored in $e$ were checked using the correct value of $e$.

Now, consider two base vertices on the same trajectory in layers $\tau$ and $\tau+\ell$, with $0\leq a_i<q$. If the trajectory is failing, choose a constraint $C$ violated by its encoded assignment. The channel corresponding to $C$ and the energy guess $z=z_1$ then passes all checks and provides a three-edge shortcut between the two vertices. Conversely, a complete dipath through the channel corresponding to $C$ and $z$ guarantees that $z_1=z$ and that the encoded assignment violates $C$.

Thus, these channels provide shortcuts precisely for failing trajectories. This construction uses $4B|\mathcal C|$ channels per dyadic interval of length at least two. \Cref{alg:lb-unoptimized} summarizes the construction.

\begin{algorithm}[!t]
\caption{Detailed graph construction}\label{alg:lb-unoptimized}
\renewcommand{\algorithmicrequire}{\textbf{Input:}}
\begin{algorithmic}[1]
\setlength{\itemsep}{0.15em}
\Require $U,L,q$, constraints $\mathcal C$, and the digit decoder
\State Create all base vertices $(\tau,\mathbf x,z)$ with $\tau=0,\ldots,L$ and $(\mathbf x,z)\in U$.
\Statex $\triangleright$ \textit{Connect vertices on the same trajectory in consecutive layers.}
\For{each unit interval $I=[\tau,\tau+1]$}
  \State \parbox[t]{\dimexpr\linewidth-\algorithmicindent\relax}{\raggedright
  Create two copies of $U$ as auxiliary layers between layers $\tau$ and $\tau+1$.}
\State \parbox[t]{\dimexpr\linewidth-\algorithmicindent\relax}{\raggedright
  For each $a_1\in[q]$, $(\mathbf x, z) \in U$,
  add $(\tau, \mathbf{x},z)\to(x_1+a_1,x_2,x_3,z+a_1^2)$.}
\State \parbox[t]{\dimexpr\linewidth-\algorithmicindent\relax}{\raggedright
  For each $a_2\in[q]$, $(\mathbf x, z) \in U$, add $(\mathbf{x},z)\to(x_1,x_2+a_2,x_3,z+a_2^2)$.}
\State \parbox[t]{\dimexpr\linewidth-\algorithmicindent\relax}{\raggedright
  For each $a_3\in[q]$, $(\mathbf x, z) \in U$, add $(\mathbf{x},z)\to(\tau+1, x_1,x_2,x_3+a_3,z+a_3^2)$.}
\EndFor
\Statex $\triangleright$ \textit{Add constant-length shortcuts across dyadic intervals for failing trajectories.}
\For{each dyadic interval $I=[\tau,\tau+\ell]$ with $\tau=j\ell$ and $\ell\geq2$}
  \For{each $C\in\mathcal C$ and integer guess $0\leq z<4B$}
    \State \parbox[t]{\dimexpr\linewidth-\algorithmicindent*2\relax}{\raggedright
      Create two copies of $U$ as auxiliary layers for each $C$ and $z$.}
    \For{each $i=1,2,3$, $a_i\in[q]$, and $(\mathbf{x},z') \in U$}
      \If{$i=1$}
        \State \parbox[t]{\dimexpr\linewidth-\algorithmicindent*4\relax}{\raggedright
          If every variable of $C$ encoded in $b_i$, $a_i$, or $e$
          equals its forbidden value, add\\[0.35ex]
          $(\tau,\mathbf x,z')\to(x_1+\ell a_1,x_2,x_3,z'+\ell a_1^2)$.}
      \ElsIf{$i=2$}
        \State \parbox[t]{\dimexpr\linewidth-\algorithmicindent*4\relax}{\raggedright
          If every variable of $C$ encoded in $b_i$ or $a_i$
          equals its forbidden value, add\\[0.35ex]
          $(\mathbf x,z')\to(x_1,x_2+\ell a_2,x_3,z'+\ell a_2^2)$.}
      \ElsIf{$i=3$}
        \State \parbox[t]{\dimexpr\linewidth-\algorithmicindent*4\relax}{\raggedright
          If every variable of $C$ encoded in $b_i$ or $a_i$
          equals its forbidden value and $z' + \ell a_3^2 = z$, add\\[0.35ex]
          $(\mathbf x,z')\to(\tau+\ell,x_1,x_2,x_3+\ell a_3,z)$.}
      \EndIf
    \EndFor
  \EndFor
\EndFor
\end{algorithmic}
\end{algorithm}

\paragraph{Specifying the encoding.}

We now specify the digit allocation and how choose ranges large enough to represent every candidate assignment. Each variable in $X_1, \dots, X_K$ takes a value in $[N]$, so one base-$N$ digit suffices to store it. For each trajectory, we write each component of $\mathbf b$ and $\mathbf a$, as well as $e$, in base $N$ and use the resulting digits to encode a unique assignment to $X_1,\ldots,X_K$.

We now choose $q$ and $L$ and specify how many variables to encode in each of these components. Since $L$ must be a power of two for the dyadic intervals, let \EMPH{$Q$} be the smallest power of two at least $2N$, so that $2N\leq Q<4N$. Choose \EMPH{$q=Q^\kappa$} and \EMPH{$L=Q^\lambda$}, where $\kappa$ and $\lambda$ are positive integer constants to be fixed in \Cref{sec:lb-parameters}. Then the bounds $2Lq$ for each $x_i$ and $4Lq^2$ for $z$ are polynomial in $N$.

Since $N^\kappa\leq q$, each direction component
$a_i$ can store at least $\kappa$ variables from $X_1, \dots, X_K$. Similarly, each intercept component $b_i$ can store $\kappa+\lambda$ variable values because $Lq=Q^{\kappa+\lambda}$. The energy coordinate $e$ in layer $0$ can store \EMPH{$s=2\kappa+\lambda$} variable values since \EMPH{$B=Q^s$}.

Thus, to store all the variables in $X_1, \dots, X_K$, we let \EMPH{$3\kappa+3(\kappa+\lambda)+s=4s= K$}. Assign $X_1,\ldots,X_K$
in order to the base-$N$ digits of $a_1,a_2,a_3,b_1,b_2,b_3,e$,
using $\kappa$ digits in each $a_i$, $\kappa+\lambda$ digits in
each $b_i$, and $s$ digits in $e$. Within each component, assign
variables starting with the least significant digit. This order is arbitrary since we only need to fix a one-to-one correspondence between the variables and these digit positions.

Given any assignment to $X_1,\ldots,X_K$, form each $a_i$, $b_i$, and $e$ by placing the values of its assigned variables in these fixed digit positions. The resulting integers satisfy $0\leq a_i<N^\kappa\leq q$,
$0\leq b_i<N^{\kappa+\lambda}\leq Lq$, and $0\leq e<N^s\leq B$. Hence, every candidate assignment is represented by a trajectory in the graph.

In the next subsection, we change how variables are encoded in $e$ by using the Chinese remainder theorem so that checking $C$ requires guessing only the energy information needed to read the variables of $C$ stored in $e$. This will further decrease the size of the constructed graph and give us a stronger reduction.

\subsection{Compressing the energy encoding with the Chinese remainder theorem}\label{sec:lb-crt}

Recall that, for each constraint $C$ and dyadic interval
$[\tau,\tau+\ell]$ with $\ell\geq2$, the previous construction creates a separate channel for every possible energy coordinate of the base vertex in layer $\tau+\ell$. This requires $4B$ guesses and two auxiliary layers for each guess. However, $C$ involves only three variables, and we only need the information that determines the values of those stored in $e$. Intuitively, if we could divide the energy coordinate encoding into parts that can be read independently, then checking $C$ would require only the parts storing its variables. We could then guess only the relevant energy information, reducing the number of auxiliary layers and the cost of constructing the graph. Here, Chinese remainder theorem comes to rescue. 

\paragraph{Encoding variables by remainders.}

For distinct primes $p_1,\ldots,p_s$, the Chinese remainder theorem states that any prescribed remainders modulo these primes determine a unique integer between $0$ and $\prod_{r=1}^s p_r-1$. Thus, we can encode an assignment to $X_1, \dots, X_r$ by prescribing remainders that represent its variable values. These remainders then uniquely determine an energy coordinate $e$ in the range $\prod_{r=1}^s p_r-1$.

Recall that the energy coordinate $e$ in layer $0$ stores $s$ of the variables $X_1,\ldots,X_K$. To change how these variables are encoded, we need the remainders to represent every possible assignment to these $s$ variables. Under these conditions, changing the encoding preserves our ability to represent every candidate assignment.

More specifically, we use one remainder to store one variable. We choose $s$ distinct primes $p_1,\ldots,p_s\in[N,2N]$ and use the remainders $e\bmod p_1,\ldots,e\bmod p_s$ to encode the variables stored in $e$, in the same order as in the previous encoding. Since each prime satisfies $p_r\geq N$, so its possible remainders include all the variable values in $[N]$.

Since $s$ is fixed, the prime number theorem guarantees that $[N,2N]$ contains at least $s$ distinct primes for all sufficiently large $N$. Recall that $Q\geq2N$ and $B=Q^s$, so these primes satisfy
$$
N^s\leq\prod_{r=1}^s p_r\leq(2N)^s\leq Q^s=B.
$$
Thus, by the Chinese remainder theorem, for any assignment to the $s$ variables stored in $e$, there exists an integer $0\leq e<\prod_{r=1}^s p_r\leq B$ such that each remainder $e\bmod p_r$ equals the prescribed value of the variable encoded by that remainder.

We keep the encodings of $\mathbf a$ and $\mathbf b$ unchanged, so the chosen components satisfy $0\leq a_i<q$ and $0\leq b_i<Lq$. For every layer $0\leq\tau\leq L$, the corresponding coordinates therefore satisfy
$$
0\leq b_i+\tau a_i<2Lq,
\qquad
0\leq e+\tau\|\mathbf a\|^2<B+3Lq^2=4B.
$$
Thus, every base vertex on this trajectory from layer $0$ to layer $L$ belongs to the graph, which means every candidate assignment to $X_1, \dots, X_r$ has such a trajectory representation.

Once the primes are fixed, we read the value of the variable assigned with respect to $p_r$ as $(e\bmod p_r)\bmod N$. The final reduction modulo $N$ ensures that the value belongs to $[N]$.

\paragraph{Checking constraints using energy remainders.}

For a constraint $C$, let \EMPH{$M_C$} be the product of all primes $p_r$ for which the remainder $e\bmod p_r$ encodes a variable of $C$. If no variable of $C$ is stored in $e$, let $M_C=1$.

Now, consider a dyadic interval $I=[\tau,\tau+\ell]$ with
$\ell\geq2$ and $\tau=j\ell$. To construct a constant-length shortcut, instead of guessing the entire energy coordinate $z_1$ in layer $\tau=j\ell$, we guess only its remainder $r_1=z_1\bmod M_C$. To see this, suppose that the energy coordinate in layer $\tau = j\ell$ is $z_0$. Then, we have  
\begin{equation}\label{eq:lb-residue-test}
     e\equiv(j+1)z_0-jz_1 \equiv (j+1)z_0-jr_1\pmod{M_C}.
\end{equation}
Thus, by guessing $r_1$, it is enough to check the variables of $C$ stored in $e$.

We now modify the construction of the constant-length shortcuts accordingly. For each $C$ and each guessed remainder $r_1\in[M_C]$, create two copies of $U$ as auxiliary layers between layers $\tau$ and $\tau+\ell$. As before, a dipath through these auxiliary layers chooses $a_1,a_2,a_3$ on its three edges, respectively. The edge choosing $a_i$ changes $x_i$ to $x_i+\ell a_i$ and $z$ to $z+\ell a_i^2$. When choosing $a_i$, we check that every variable of $C$ stored in $a_i$ or $b_i=x_i-\tau a_i$ equals its forbidden value.

On the edge from layer $\tau$ to the first auxiliary layer, use $((j+1)z_0-jr_1)\bmod M_C$ as the to read the variables of $C$ stored in $e$. Add the edge whenever no variable of $C$ stored in $a_1$, $b_1$, and $e$ has a value different from its forbidden value. Add an edge from the first auxiliary layer to the second auxiliary layer if no variable of $C$ stored in $a_2$ or $b_2$ differs from its forbidden value. Add an edge from the second auxiliary layer to a base vertex in layer $\tau+\ell$ if no variable of $C$ stored in $a_3$ or $b_3$ differs from its forbidden value and the energy coordinate $z_1$ of that base vertex satisfies $z_1\bmod M_C=r_1$.

\paragraph{New construction analysis.}

Fix a constraint $C \in \mathcal{C}$, we now only need $M_C$ pairs of auxiliary layers instead of $4B$. Since $C$ involves only three variables, $M_C$ is a product of at most three primes. Hence, $M_C\leq(2N)^3$. Since $|\mathcal C|\leq\binom K3N^3$, we have $\sum_{C\in\mathcal C}M_C=O_K(N^6)$.

\Cref{alg:lb-crt} summarizes the changes to the shortcut construction in~\Cref{alg:lb-unoptimized}. Together with the edges connecting consecutive layers from~\Cref{alg:lb-unoptimized}, these shortcuts give the final graph.

\begin{algorithm}[H]
\caption{Shortcut construction using energy remainders}\label{alg:lb-crt}
\renewcommand{\algorithmicindent}{1.25em}
\begin{algorithmic}[1]
\setlength{\itemsep}{0.3em}
\State Choose $s$ distinct primes $p_1,\ldots,p_s\in[N,2N]$.
\Statex $\triangleright$ \textit{Add constant-length shortcuts across dyadic intervals for failing trajectories.}
\For{each dyadic interval $I=[\tau,\tau+\ell]$ with $\tau=j\ell$ and $\ell\geq2$}
  \For{each $C\in\mathcal C$ and guessed remainder $r_1\in[M_C]$}
    \State \parbox[t]{\dimexpr\linewidth-\algorithmicindent*2\relax}{\raggedright
      Create two copies of $U$ as auxiliary layers for $C,r_1$
      between layers $\tau$ and $\tau+\ell$.}
    \For{each $i=1,2,3$, $a_i\in[q]$, and $(\mathbf x,z')\in U$}
      \State Let $b_i=x_i-\tau a_i$.
      \If{$i=1$}
        \State \parbox[t]{\dimexpr\linewidth-\algorithmicindent*4\relax}{\raggedright
          If no variable of $C$ encoded in $b_i$, $a_i$, or\\
          $(((j+1)z'-jr_1)\bmod p_r)$ differs from its forbidden value, add\\[0.6ex]
          \hspace*{\algorithmicindent}
          $(\tau,\mathbf x,z')\to(x_1+\ell a_1,x_2,x_3,z'+\ell a_1^2)$.}
      \ElsIf{$i=2$}
        \State \parbox[t]{\dimexpr\linewidth-\algorithmicindent*4\relax}{\raggedright
          If no variable of $C$ encoded in $b_i$ or $a_i$
          differs from its forbidden value, add\\[0.6ex]
          \hspace*{\algorithmicindent}
          $(\mathbf x,z')\to(x_1,x_2+\ell a_2,x_3,z'+\ell a_2^2)$.}
      \ElsIf{$i=3$}
        \State \parbox[t]{\dimexpr\linewidth-\algorithmicindent*4\relax}{\raggedright
          If $(z'+\ell a_3^2)\bmod M_C=r_1$ and no variable of $C$
          encoded in $b_i$ or $a_i$ differs from its forbidden value, add\\[0.6ex]
          \hspace*{\algorithmicindent}
          $(\mathbf x,z')\to(\tau+\ell,x_1,x_2,x_3+\ell a_3,z'+\ell a_3^2)$.}
      \EndIf
    \EndFor
  \EndFor
\EndFor
\end{algorithmic}
\end{algorithm}

\subsection{The diameter gap}\label{sec:lb-gap}

Recall that every candidate assignment is represented by a trajectory,
and that we add shortcuts only for failing trajectories. We now show
that these properties give the required diameter gap. If a satisfying
assignment exists, we will find two base vertices at distance $3L$.
If every assignment violates a constraint, we will first bound the distance between
vertices on the same trajectory, and then extend the bound to any
two vertices connected by a dipath.

We begin by applying the argument from \Cref{sec:lb-trajectories}
to the graph constructed in \Cref{sec:lb-unoptimized,sec:lb-crt}.

\begin{lemma}[Dipaths between vertices on the same trajectory]\label{lem:lb-rigidity}
Let $u$ and $v$ be a valid pair on the trajectory with direction
$\mathbf a$, with $u$ in an earlier layer than $v$.
Every dipath from $u$ to $v$ always follows direction $\mathbf a$ when moving through consecutive layers and or using a shortcut.
\end{lemma}
\begin{proof}
Consider any dipath from $u$ to $v$. Between successive base vertices, it uses either a three-edge dipath connecting consecutive layers or a three-edge shortcut across a dyadic interval.

For the $j$-th pair of successive base vertices, let $\mathbf a_j$
be the direction chosen by the three edges between them, and let
$\ell_j$ be the difference between their layers. The three edges together increase $\mathbf x$ by $\ell_j\mathbf a_j$ and $z$ by
$\ell_j\|\mathbf a_j\|^2$. Thus, \eqref{eq:lb-rigidity} gives $\mathbf a_j=\mathbf a$ for every $j$.

\end{proof}

\begin{lemma}[Diameter gap]\label{lem:lb-gap}
Let $D$ denote the reachability diameter of the constructed digraph by \Cref{alg:lb-unoptimized} and \Cref{alg:lb-crt}.  If there is an assignment to $X_1,\ldots,X_K$ satisfying every
constraint in $\mathcal C$, then $D\geq3L$. Otherwise,
\begin{equation}\label{eq:lb-short-diameter}
                  D\leq \EMPH{D_{\mathrm{NO}}}:=30\log L\,\log(4B)+6.
\end{equation}
\end{lemma}
\begin{proof}
\emph{When a satisfying assignment exists.}
By \Cref{sec:lb-crt}, the assignment is represented by a trajectory
whose base vertices in every layer from $0$ to $L$ belong to the graph.
Let $u$ and $v$ be its base vertices in layers $0$ and $L$.
The construction connects its vertices in consecutive layers by
three-edge dipaths, giving a dipath of length $3L$ from $u$ to $v$.

By \Cref{lem:lb-rigidity}, every dipath from $u$ to $v$ follows
the same direction, and all its base vertices lie on this satisfying
trajectory. Such a dipath cannot use a shortcut since the shortcuts are only added between valid pairs on a failing trajectory. Hence, every dipath from
$u$ to $v$ uses only connections between consecutive layers.
Each such connection has three edges, so $ D\geq d(u,v)=3L$.

\emph{When there is no satisfying assignment.}
Suppose now that every assignment violates a constraint in $\mathcal C$. We first consider two base vertices on the same trajectory, with $a_i\in[q]$ for $i=1,2,3$.

Since every assignment violates a constraint, this trajectory is failing. Consider a dyadic interval $[\tau,\tau+\ell]$ between the layers of the two fixed base vertices. If $\ell\geq2$, choose a constraint $C$ violated by the trajectory's assignment and
$r_1=z_1\bmod M_C$, where $z_1$ is the energy coordinate of its base vertex in layer $\tau+\ell$. By \eqref{eq:lb-residue-test}, the first edge reads the correct values of the variables of $C$
stored in $e$. All the variable checks pass, and the last edge verifies $r_1$. Hence, the two auxiliary layers for $C,r_1$ give a three-edge shortcut. 

As explained in \Cref{sec:lb-trajectories}, the interval between the layers of the two fixed vertices can be partitioned into at most $2\log L$ dyadic intervals.  Thus, concatenating the three-edge dipaths for these intervals gives a dipath of length at most $6\log L$.

Nevertheless, two base vertices connected by a dipath may not be a valid pair and they do not determine any trajectory. In this case, we can divide the dipath into segments of constant direction, but there may be too many segments to obtain a short dipath by replacing each segment separately. Instead, there exists a dipath using only a logarithmic number of directions.
Thus, there exists a dipath following each direction for consecutive layers and the $O(\log L)$ bound be applied to each resulting segment.

To bound the number of directions, we apply the nonnegative
integer-cone bound of Eisenbrand and
Shmonin~\cite[Theorem~1(i)]{eisenbrandshmonin2006cones},
using the coordinatewise bound in the proof
of~\cite[Theorem~5.1]{shmonin2007thesis}.

Let $u=(\tau_0,\mathbf x_0,z_0)$ and
$v=(\tau_1,\mathbf x_1,z_1)$ be distinct base vertices with
a dipath from $u$ to $v$. Suppose that there is a dipath from $u$ to $v$ that can be divided into segments of distinct direction $\mathbf a_1,\ldots,\mathbf a_r$. Thus, we have 
$$
 (\tau_1-\tau_0,\mathbf x_1-\mathbf x_0,z_1-z_0)
   =\sum_{j=1}^r\ell_j(1,\mathbf a_j,\|\mathbf a_j\|^2),
 \qquad \ell_j\in\mathbb Z_{>0}.
$$

\begingroup
\clubpenalty=10000
\displaywidowpenalty=10000
By the nonnegative integer-cone bound of Eisenbrand and Shmonin~\cite[Theorem~1(i)]{eisenbrandshmonin2006cones}, using the coordinatewise estimate in the proof of~\cite[Theorem~5.1]{shmonin2007thesis}, we can choose the representation so that
$$
r\leq \log(\tau_1-\tau_0+1)
  +\sum_{i=1}^3\log(x_{1,i}-x_{0,i}+1)
  +\log(z_1-z_0+1),
$$
where $x_{0,i}$ and $x_{1,i}$ are the $i$-th coordinates of $\mathbf x_0$ and $\mathbf x_1$, respectively. Since $\tau_1-\tau_0\leq L$, $x_{1,i}-x_{0,i}<2Lq$,
and $z_1-z_0<4B$, and all these differences are integers,
\begin{equation}\label{eq:lb-few-directions}
  r\leq \log(L+1)+3\log(2Lq)+\log(4B)
    \leq 5\log(4B).
\end{equation}
\par
\endgroup

Starting at $u$, advance by $\ell_j$ layers in direction
$\mathbf a_j$ for $j=1,\ldots,r$, in order. The total changes
reach $v$, and all coordinates are nondecreasing. Hence, the
intermediate coordinates stay between those of $u$ and $v$,
within $U$, so the construction provides the required dipath.
By the preceding argument, the distance of each segment is at most $6\log L$. Concatenating these short dipaths gives $d(u,v)\leq6r\log L\leq30\log L\,\log(4B) = D_{NO}$.

Finally, if a dipath contains base vertices, the portions before
its first base vertex and after its last base vertex each have
at most three edges. Replacing the dipath between these base
vertices by the short dipath above adds at most six edges to the
bound. A dipath with no base vertex stays within one pair of
auxiliary layers and has at most three edges. This proves
\eqref{eq:lb-short-diameter}.
\end{proof}

\subsection{Graph size and construction time}\label{sec:lb-size}

We now count the vertices and edges created by
\Cref{alg:lb-unoptimized,alg:lb-crt}. Recall that each base or
auxiliary layer contains $|U|$ vertices. There are $L+1$ base
layers and two auxiliary layers for each of the $L$ pairs of
consecutive layers. For each of the $L-1$ dyadic intervals of
length at least two, we create two auxiliary layers for every constraint
$C\in\mathcal C$ and guessed remainder $r_1\in[M_C]$. Thus,
shortcuts across each such interval require
$\sum_{C\in\mathcal C}M_C$ pairs of auxiliary layers.

For each pair of auxiliary layers, we add edges from the earlier
base layer to the first auxiliary layer, from the first to the
second auxiliary layer, and from the second auxiliary layer to
the later base layer. For each of these three connections between
layers, each vertex in the earlier layer has at most $q$ outgoing
edges, one for each choice of the corresponding direction
component. Thus, each pair of auxiliary layers contributes at
most $3|U|q$ edges. Therefore,
\begin{align}
 n&=(3L+1)|U|+2(L-1)|U|\sum_{C\in\mathcal C}M_C,\label{eq:lb-size}\\
 m&\leq3|U|q\Bigl[L+(L-1)\sum_{C\in\mathcal C}M_C\Bigr].\nonumber
\end{align}

To construct the graph, we first list the constraints in
$\mathcal C$ in $O_K(N^3)$ time and find the required primes in
$[N,2N]$ by a sieve in $\widetilde O(N)$ time. We then create
all base and auxiliary vertices and consider the choices of
$a_i\in[q]$ for each set of edges described above. For each
choice, we compute the new coordinates and perform the variable
and remainder checks from \Cref{sec:lb-unoptimized,sec:lb-crt}.
Since $\kappa$ and $\lambda$ are fixed, all integers involved
have $O_K(\log N)$ bits, so each choice takes at most
polylogarithmic time. The total construction time is therefore
\begin{equation}\label{eq:lb-construction-time}
 \widetilde O_K\Bigl(|U|qL\Bigl(1+\sum_{C\in\mathcal C}M_C\Bigr)+N^3\Bigr).
\end{equation}

Recall that $Q=\Theta(N)$, $L=Q^\lambda$, $q=Q^\kappa$, and
$|U|=32L^4q^5$. Substituting these values and
$\sum_{C\in\mathcal C}M_C=O_K(N^6)$ from \Cref{sec:lb-crt}
gives
\begin{equation}\label{eq:lb-general-exponents}
 \begin{aligned}
 n&=O_K\bigl(N^{5\lambda+5\kappa+6}\bigr),\\
 m&=O_K\bigl(N^{5\lambda+6\kappa+6}\bigr),\\
 \text{construction time}&=\widetilde O_K\bigl(N^{5\lambda+6\kappa+6}\bigr).
 \end{aligned}
\end{equation}

These bounds explain why we changed the energy encoding in
\Cref{sec:lb-crt}. Guessing the entire energy coordinate would
require $4B|\mathcal C|=O_K(N^{2\kappa+\lambda+3})$ pairs of
auxiliary layers per dyadic interval of length at least two.
The resulting construction-time bound would have exponent
$8\kappa+6\lambda+3$, exceeding $K=8\kappa+4\lambda$.
Guessing only the required remainder reduces the number of pairs
to $O_K(N^6)$, whose exponent does not grow with the number of
variables stored in $e$.

\subsection{Choosing the parameters and obtaining the exponent}\label{sec:lb-parameters}

We now choose $\kappa$ and $\lambda$ so that the reduction
satisfies two requirements. First, constructing the graph and
running a $\widetilde O(n+m)$-time approximation algorithm must
take $O(N^{K-\eta})$ time for some constant $\eta>0$.
Second, the approximation guarantee must distinguish the case
where a satisfying assignment exists from the case where every
assignment violates a constraint.

Recall that the construction encodes $K=8\kappa+4\lambda$
variables. By \eqref{eq:lb-general-exponents}, constructing the
graph and running the approximation algorithm together take
$\widetilde O_K(N^{5\lambda+6\kappa+6})$ time. Thus, the first
requirement holds whenever $5\lambda+6\kappa+6<K$, or
equivalently, $\lambda<2\kappa-6$.

For the second requirement, consider an $n^\beta$-approximation
for a constant $\beta>0$. By \Cref{lem:lb-gap}, if a satisfying
assignment exists, the estimate is at least $3L/n^\beta$;
otherwise, it is at most $D_{\mathrm{NO}}$. We therefore need
$3L/n^\beta>D_{\mathrm{NO}}$. Since
$D_{\mathrm{NO}}=O_K((\log N)^2)$, $L=\Theta_K(N^\lambda)$,
and $n=O_K(N^{5\lambda+5\kappa+6})$, it suffices to choose
\begin{equation}\label{eq:lb-gap-budget}
                 \beta<\frac{\lambda}{5\lambda+5\kappa+6}.
\end{equation}
For fixed $\kappa$, this ratio increases with $\lambda$.
We therefore take $\lambda=2\kappa-7$, the largest integer
satisfying $\lambda<2\kappa-6$, and choose $\kappa\geq4$ so
that $\lambda$ is positive. Then $K=16\kappa-28$, and
\eqref{eq:lb-general-exponents} becomes
\begin{equation}\label{eq:lb-final-size}
 \begin{aligned}
 n&=O_K(N^{15\kappa-29}),\\
 m&=O_K(N^{16\kappa-29})=O_K(N^{K-1}),\\
 \text{construction time}&=\widetilde O_K(N^{K-1}).
 \end{aligned}
\end{equation}
At the same time, $L=\Theta_K(N^{2\kappa-7})$. For this
choice of $\lambda$, the right-hand side of
\eqref{eq:lb-gap-budget} satisfies
\begin{equation}\label{eq:lb-beta}
 \frac{2\kappa-7}{15\kappa-29}
   =\frac{2}{15}-\frac{47}{15(15\kappa-29)}
   \longrightarrow\frac{2}{15}
 \qquad\text{as }\kappa\to\infty.
\end{equation}
Thus, for every fixed $\varepsilon\in(0,2/15)$, we can choose
$\kappa$ large enough that $\beta=2/15-\varepsilon$ satisfies
both requirements. This gives the exponent in \Cref{thm:lb-main}.

\paragraph{Optimizing the number of coordinates.}
We can now justify the choice of three coordinates for
$\mathbf x$ in \Cref{sec:lb-trajectories}. Suppose instead that
$\mathbf x$ has $d$ coordinates for a fixed integer $d\geq2$.
Both a connection between consecutive layers and a shortcut
would use $d-1$ auxiliary layers, with the $d$ edges choosing
$a_1,\ldots,a_d$ separately. Each $a_i$ would
store $\kappa$ variables, each $b_i$ would store
$\kappa+\lambda$ variables, and $e$ would still store
$2\kappa+\lambda$ variables. Thus, a trajectory would encode
$K=(d+1)(2\kappa+\lambda)$ variables.

Each $x_i$ would still range over $[2Lq]$. Since the energy
increase across $L$ layers is less than $dLq^2$, we would let
$z$ range over $[(d+1)B]$. Each base or auxiliary layer would
then contain $O_d(L^{d+1}q^{d+2})$ vertices. The same counting
argument therefore gives a vertex bound of
$O_K(N^{(d+2)(\kappa+\lambda)+6})$ and a construction-time bound of
$\widetilde O_K(N^{(d+3)\kappa+(d+2)\lambda+6})$.

To keep the construction time polynomially smaller than $N^K$,
we would need $\lambda<(d-1)\kappa-6$. For fixed $d$, the
same diameter analysis gives a lower bound of $dL$ when a
satisfying assignment exists and an upper bound of
$O_K((\log N)^2)$ otherwise. Taking $\lambda=(d-1)\kappa-7$
and letting $\kappa$ grow therefore gives the
limiting approximation exponent $(d-1)/(d(d+2))$.
For every integer $d\geq2$, this is at most $2/15$, with
equality only at $d=3$: the inequality is equivalent to
$(2d-5)(d-3)\geq0$, which holds for $d=2$ and for all
$d\geq3$. Thus, three coordinates give the largest exponent
for this construction under the size bounds above.

%% file: UpperBound.tex
\section{Upper Bound}\label{sec:ReachDiam}

\subsection{Algorithm}

\begin{algorithm}[!ht]
    \caption{$\textsc{ReachDiam}(G)$.}
    \label{alg:ReachDiam}
    \textbf{Input:} Digraph $G$ with $n$ vertices.\par
    \textbf{Output:} Estimate $\widehat{D}$ to the reachability diameter $D$ of $G$. \par
    \medskip
    \textbf{Procedure:} $\textsc{ReachDiam}(G).$
    \begin{enumerate}
        \setlength{\itemsep}{2pt}
        \setlength{\parsep}{0pt}
        \item Base case: if $|E(G)|=0$, return 0.
        \item Repeat \Call{Recurse}{G, 0} for $O(\log n)$ trials, 
        throwing out each trial exceeding $c_1(m+n)\log^{c_2}(n+2)$ computation steps,
        where $c_1, c_2$ are large enough constants.
        \item Return the maximum of 1 and the values reported by \textsc{Recurse}.
    \end{enumerate}
    \medskip
    \textbf{Procedure:} $\textsc{Recurse}(H, r)$. \par
    Here $r$ denotes the recursion level.
    \begin{enumerate}
        \setlength{\itemsep}{2pt}
        \setlength{\parsep}{0pt}
        \item Base case: if $|E(H)|=0$, return 0.
        \item \Call{Sweep}{H}. \steplabel{step:sweep}
        \item Sample $S \subseteq V(H)$ independently with probability: 
        \[
        p_r~\gets~\min\{1,20\Theta(\log{n})^{r+1}\log(n)/n\}.
        \]
        \item For each $s \in S$, run forward and backward BFS starting at $s$ in the original $H$. The vertices $R_H^+(s)\cap R_H^-(s)$ seen in both searches are omitted in subsequent steps. \steplabel{step:BFS} 
        \item Group the remaining vertices into sets $V_1, \dots$ according to their reachability signatures:
        \[
            \sigma(v) := \bigl(\mathbf{1}[v\leadsto s], \mathbf{1}[s\leadsto v] \bigr)_{s\in S}.
        \]
        \item For each $V_i$, \Call{Recurse}{$H[V_i]$, $r+1$}.
        \item Return the maximum of 1, the value reported by $\Call{Sweep}{}$, the largest depth of any search, and the values reported by recursive calls. \steplabel{step:reporting}
    \end{enumerate}
\end{algorithm}

\paragraph{Running time.}

Each call $\Call{Sweep}{H}$ takes $O(V(H) + E(H))$ time. 
Recursive inputs at each of $O(\log n)$ levels are vertex-disjoint, 
so the total running time of $\Call{Sweep}{}$ is $O((m+n)\log n)$.
Together with Proposition 4.1 of \cite{ashvinkumar2026parallel}, this implies that, for sufficiently large constants $c_1, c_2$, an uncapped trial including the subroutine $\Call{Sweep}{}$ finishes within $c_1(m+n)\log^{c_2}(n)$ steps with probability $1-O(n^{-10})$.
The worst case running time is $\widetilde{O}(m+n)$.

\subsection{Correctness} 
This section is devoted to proving the correctness of \Cref{alg:ReachDiam}.
For the sake of analysis, fix $P$ an arbitrary diametral shortest path.
Consider the recursion tree $\mathcal{T}$ of an uncapped trial of \Cref{alg:ReachDiam}, whose nodes are recursive instances and whose edges connect each child to the parent that called it.
Let $\mathcal{T}_P$ be the subtree of $\mathcal{T}$ induced by the recursive instances $H$ for which $P[V(H)] \neq \emptyset$ -- by \Cref{lem:contiguous-subpath} each such instance contains a contiguous subpath of $P$.
All content below assumes that $(H, P^\prime = P[V(H)])$ is a node of $\mathcal{T}_P$, where $P^\prime$ is the non-empty, contiguous subpath of $P$ contained entirely in $H$.

Recall that a vertex $v$ is called a \textbf{bridge} of a path $Q$ when $v \rightsquigarrow Q$ and $Q\rightsquigarrow v$.
We call a node $(H, P^\prime)$ of $\mathcal{T}_P$ \textbf{bridged}, if a sampled pivot is a bridge of $P^\prime$. 
Furthermore, a node $(H, P^\prime)$ is \textbf{small}, if $|R_H(P^\prime)| \leq D$, and \textbf{large} otherwise.

Now, starting at the root prune $\mathcal{T}_P$ in a depth-first manner, stopping each branch at the first small or bridged node (hereafter we use $\mathcal{T}_P$ to refer to the pruned tree, otherwise known as the subproblem tree). 
Along with \Cref{obs:bridge-metric}, the following lemma establishes the mechanisms by which small and bridged nodes of $\mathcal{T}_P$ reveal the length of the subpath of $P$ they contain.
\begin{lemma}\label{lem:sweep-metric}
    If $(H, P^\prime=p_1, \dots, p_{|P^\prime|})$ is small, then with probability at least $|V(P^\prime)|/(2D)$, \Cref{alg:ReachDiam} observes an original distance in the base input graph at least $d(P^\prime)/2$ during the \textup{$\Call{Sweep}{}$} phase.
\end{lemma}
\begin{proof}
    Recall that the least ancestor $a$ of $P^\prime$ is the first uncolored member of $P^\prime$'s in-relevance $R_H^-(P^\prime)$ encountered during the execution of $\Call{Sweep}{}$.
    By contradiction with $a$'s minimality,
    both $a$ and $P^\prime$ must be uncolored at the moment that $a$ is encountered,
    otherwise they would have been discovered by search from an ancestor earlier in the execution.
    Thus, 
    if $a \in \{p_1, \dots, p_{\lceil |P^\prime|/2 \rceil}\}$, 
    then $\Call{Sweep}{}$ observes $d(a, p_{|P^\prime|}) \geq d(P^\prime)/2$ during BFS rooted at $a$.
    By Propositions 2.1-2.2 of \cite{cohen1997size},
    the least ancestor $a$ of $P^\prime$ is distributed uniformly at random on $R_H^-(P^\prime)$.
    Therefore, the probability that $a \in \{p_1, \dots, p_{\lceil |\mathcal{P^\prime}|/2 \rceil}\}$ is exactly:
    \[
    \frac{\lceil|P'|/2\rceil}{|R_H^-(P^\prime)|} \geq \frac{|V(P')|}{2|R_H(P^\prime)|}\geq \frac{|V(P')|}{2D} \geq \frac{d(P')}{2D},
    \]
    where we use the fact that $(H, P^\prime)$ is small to bound $R_H^-(P^\prime) \leq R_H(P^\prime) \leq D$.
    To complete the proof we show that every distance reported by $\Call{Sweep}{}$ is a real distance in the input graph,
    Consider that after each search the colored set of $\Call{Sweep}{}$ is closed under reachability by definition,
    and by consequence the uncolored set contains every dipath in $H$ between every ordered pair of remaining endpoints.
    In other words, 
    the value reported by $\Call{Sweep}{}$ is an original distance in $H$,
    which is an original distance in the base input graph $G$ by \Cref{lem:contiguous-subpath}.
\end{proof}

Thus, all we need is a small or bridged node which contains a significant length of the original diametral path $P$ 
(provided that the probability guarantee is justified). 
Vertices of $P$ are only removed from the recursion in the children of bridged nodes (\Cref{step:BFS}), 
so the set of leaves $\mathcal{F} \subseteq V(\mathcal{T}_P)$ partitions $V(P)$.
In particular, $d(P)=\sum_{i=1}^{|\mathcal{F}|}d(P^\prime_i)+(|\mathcal{F}|-1)$, where $\mathcal{F} = \{(H_1, P^\prime_1), \dots\}$.
We call $\mathcal{F}$ the \textbf{frontier}.
The analysis of \cite{ashvinkumar2026parallel} bounds the cardinality of the frontier.
\begin{lemma}\label{lem:frontier-bound}
    $|\mathcal{F}| \leq 1+ O(\log^2{n})\frac{n^{1/2+o(1)}}{\sqrt{D}}$, with probability at least $3/4$.
\end{lemma}
\begin{proof}
    Let $X$ be the random variable counting the number of large nodes in $\mathcal{T}_P$. 
    By \Cref{lem:large-node-bound-fanout-bound} with $x=D$,
    \[
    \mathbb{E}[X] \leq \frac{n^{1/2+o(1)}}{\sqrt{D}},
    \]
    and $\Pr[X \leq 8\cdot\frac{n^{1/2+o(1)}}{\sqrt{D}}] \geq 1-1/8$ by Markov.
    
    By construction, every non-root frontier node has a large parent in $\mathcal{T}_P$.
    Moreover, by the second guarantee of \Cref{lem:large-node-bound-fanout-bound}, every large node has $O(\log^2{n})$ children in $\mathcal{T}_P$ with probability at least $1-O(n^{-10})$.
    Thus:
    \[
    |\mathcal{F}| \leq 1+ 8\cdot O(\log^2{n})\frac{n^{1/2+o(1)}}{\sqrt{D}},
    \]
    with probability at least $1-1/8-O(n^{-10})\geq 3/4$, as desired.
\end{proof}

We are now ready to complete the proof of \Cref{thm:main-theorem}.
\begin{proof}
    Assume $D\geq 1$, since the case $D=0$ is trivial. 
    The frontier partitions the diametral shortest path $P$ into contiguous subpaths, separated by $|\mathcal{F}|-1$ connecting edges. 
    Thus, when $D$ is sufficiently larger than $|\mathcal{F}|$, some frontier subpath must have length $\Omega(D/|\mathcal{F}|)$. 
    If such a subpath belongs to a bridged node, \Cref{obs:bridge-metric} guarantees that the algorithm observes at least half its length, 
    which gives the desired estimate.
    
    For a small node, \Cref{lem:sweep-metric} gives a probability of observing half its subpath length that is proportional to the length of its subpath. A single long subpath may have a small success probability. However, if we have sufficiently many small nodes containing subpaths that are long enough, they collectively give a constant probability of success. We establish this by showing that short subpaths and connecting edges cannot account for too much of $P$.
    
    More precisely, condition on the pivot samples, $\mathcal{F}$ is fixed while the sweeps remain independent. If $D\leq 4|\mathcal{F}|$, the algorithm already reports at least $1\geq D/(4|\mathcal{F}|)$. We then assume $D>4|\mathcal{F}|$, so that the connecting edges contribute less than $D/4$.
    
    Next, we choose a target $t$ so that the subpaths of length at most $2t$ contribute at most another $D/4$ in total. Specifically, set $t=D/(8|\mathcal{F}|)$, and call a frontier node $(H,P')$ \textbf{short} if $d(P')\leq 2t$, and \textbf{long} otherwise. The short nodes contain total path length at most $2t|\mathcal{F}|=D/4$. Since $\sum_{(H,P')\in\mathcal{F}}d(P')=D-(|\mathcal{F}|-1)$, the long nodes contain total path length at least:
    \[
        \sum_{\substack{(H,P')\in\mathcal{F}\\ d(P')>2t}} d(P')
        \geq D-(|\mathcal{F}|-1)-2t|\mathcal{F}|
        \geq \frac{D}{2}.
    \]
    
    If any long node is bridged, \Cref{obs:bridge-metric} guarantees an observed distance of at least $d(P')/2>t$. Otherwise, every long node is small. For each such node, \Cref{lem:sweep-metric} guarantees that its sweep observes a distance at least $d(P')/2>t$ with probability at least $|V(P')|/(2D)\geq d(P')/(2D)$. The probability that all these sweeps fail to observe a distance at least $t$ is at most:
    \begin{align*}
        \prod_{\substack{(H,P')\in\mathcal{F}\\ d(P')>2t}}
            \left(1-\frac{d(P')}{2D}\right)
        &\leq
        \exp\left(
            -\frac{1}{2D}
            \sum_{\substack{(H,P')\in\mathcal{F}\\ d(P')>2t}} d(P')
        \right)\\
        &\leq e^{-1/4}.
    \end{align*}
    Thus, for every fixed choice of pivot samples, the algorithm reports at least $D/(8|\mathcal{F}|)$ with probability at least $1-e^{-1/4}$.
    
    By \Cref{lem:frontier-bound}, with probability at least $3/4$ over the pivot samples, the size of the frontier is bounded by: $|\mathcal{F}|\leq 1+O(\log^2 n)n^{1/2+o(1)}/\sqrt{D}$. 
    Combining these bounds, a single execution reports a distance at least:
    \[
        \frac{D}{
            8\left(1+O(\log^2 n)
            \frac{n^{1/2+o(1)}}{\sqrt{D}}\right)}
    \]
    with probability at least $\frac34(1-e^{-1/4})$.
    Since a trial is stopped by exceeding its computation budget with probability
    $O(n^{-10})$, a capped trial succeeds with probability at least
    $\frac34(1-e^{-1/4})-O(n^{-10})$, which remains bounded below
    by a positive constant.
    Thus, after $O(\log n)$ independent repetitions, taking the maximum reported value gives an estimate $\widehat D$ satisfying the same lower bound with high probability.
    
    If $D\leq n^{1/3}$, the bound $\widehat D\geq 1$ immediately gives $D/\widehat D\leq n^{1/3}$. Otherwise, $D>n^{1/3}$, and the preceding bound gives:
    \[
        \frac{D}{\widehat D}
        \leq
        8\left(1+O(\log^2 n)
        \frac{n^{1/2+o(1)}}{\sqrt{D}}\right)
        \leq n^{1/3+o(1)}.
    \]
    
    Moreover, \Cref{lem:contiguous-subpath} guarantees that every reported distance is a distance in the base graph, so $\widehat D\leq D$, which completes the proof.
\end{proof}